%% file: paper.tex
\documentclass[epj,nopacs,fleqn]{svjour}
\usepackage{epsfig,amsmath,amssymb}
\usepackage{graphicx, subfigure}
\usepackage{slashed}
\usepackage{xcolor}

\newcommand{\eps}{\epsilon}
\newcommand{\lam}{\lambda}

\newcommand{\Mpl}{\overline{M}_{\rm Pl}}
\newcommand{\gld}{\tilde G}

\newcommand{\go}{\tilde g}

\newcommand{\gfive}{\gamma^5}
\def\be{\begin{equation}}
\def\ee{\end{equation}}
\newcommand{\e}{\varepsilon}
\newcommand{\ebar}{{\bar\varepsilon}}
\newcommand{\sibar}{{\bar\sigma}}
\newcommand{\lambar}{{\bar\lambda}}
\newcommand{\psibar}{{\bar\psi}}
\newcommand{\Psibar}{{\bar\Psi}}
\newcommand{\alphadot}{{\dot\alpha}}
\newcommand{\Jbar}{{\bar J}}
\def\del{\partial}
\def\bsp#1\esp{\begin{split}#1\end{split}}
\def\bpm{\begin{pmatrix}} 
\def\epm{\end{pmatrix}} 

\newcommand{\oh}{\frac{1}{2}}
\newcommand{\tha}{\frac{3}{2}}

\newcommand{\dth}[2]{\left|d^\frac{3}{2}_{\mathtt{#1 #2}}(\theta)\right|^2}

\definecolor{darkgreen}{rgb}{0.0, 0.45, 0.0}

\def\d{{\rm d}}
\newcommand{\ie}{{\it i.e.}}
\newcommand{\eg}{{\it e.g.}}

\begin{document}

\title{Simulating spin-$\mathbf{\tha}$ particles at colliders}

\author{
 N.D.~Christensen\inst{1},
 P.~de Aquino\inst{2},
 N.~Deutschmann\inst{3},
 C.~Duhr\inst{4,5},
 B.~Fuks\inst{6,7},
 C.~Garcia-Cely\inst{8},
 O.~Mattelaer\inst{9,10},
 K.~Mawatari\inst{2}, 
 B.~Oexl\inst{2},
 Y.~Takaesu\inst{11}
}
\institute{ 
 PITTsburgh Particle physics, Astrophysics and Cosmology Center (PITT
 PACC),\\University of Pittsburgh, Pittsburgh, PA, 15260, USA
 \and
 Theoretische Natuurkunde and IIHE/ELEM, Vrije Universiteit Brussel,\\
 and International Solvay Institutes,
 Pleinlaan 2, B-1050 Brussels, Belgium
 \and
 Universit\'e de Lyon, F-69622 Lyon, France, 
 Universit\'e Lyon 1, Villeurbanne, CNRS/IN2P3, UMR5822,\\ 
 Institut de Physique Nucl\'eaire de Lyon, F-69622 Villeurbanne Cedex, France
 \and
 Institute for Theoretical Physics, ETH Zurich, 8093 Zurich, Switzerland
 \and
 Institute for Particle Physics Phenomenology, University of Durham, Durham, DH1 3LE, U.K.
 \and
 Theory Division, Physics Department, CERN, CH-1211 Geneva 23, Switzerland
 \and
 Institut Pluridisciplinaire Hubert Curien/D\'epartement Recherches
    Subatomiques, Universit\'e de Strasbourg/CNRS-IN2P3,
    23 Rue du Loess,  F-67037  Strasbourg, France\
 \and
 Physik-Department T30d, Technische Universit\"at M\"unchen, 
 James-Franck-Stra{\ss}e, 85748 Garching, Germany.
 \and
 Centre for Cosmology, Particle Physics and Phenomenology (CP3),\\
 Universit\'e Catholique de Louvain, B-1348 Louvain-la-Neuve, Belgium
 \and
 Department of Physics, University of Illinois at Urbana-Champaign, Urbana, IL 61801, USA
 \and
 Department of Physics and Astronomy, Seoul National University, Seoul
 151-742, Korea, \\
 and School of Physics, Korea Institute for Advanced Study (KIAS), Seoul 130-722, Korea
 }
\date{Received: date / Accepted: date}

\abstract{
  Support for interactions of spin-$\tha$ particles is implemented in the
  {\sc FeynRules} and {\sc ALOHA} packages and tested with the {\sc MadGraph}~5 and
  {\sc CalcHEP} event generators in the context of three phenomenological applications. 
   In the first, we implement a spin-$\tha$ Majorana gravitino field, as in local
  supersymmetric models, and study gravitino and gluino pair-production. 
  In the second, a spin-$\tha$ Dirac top-quark excitation, inspired from
  compositness models, is implemented. 
  We then investigate both top-quark excitation and
  top-quark pair-production. In the third, a general effective operator for a spin-$\tha$
  Dirac quark excitation is implemented, followed by a calculation of the angular distribution of
  the $s$-channel production mechanism.
}

\titlerunning{Simulating spin-$\tha$ particles at colliders}
\authorrunning{N.D.~Christensen \textit{et al.}}

\maketitle


\vspace*{-16.5cm}
\noindent
\small{CERN-PH-TH/2013-187, IPPP/13/56, DCPT/13/172,\\
       MCNET-13-09, PITT-PACC-1307, KIAS-P13044,\\
       TUM-HEP 900/13}\\
\today
\vspace*{13.4cm}


\section{Introduction}\label{sec:intro}

\input{intro.tex}

\section{Spin-$\mathbf{\tha}$ implementation}\label{sec:implement}

\subsection{{\sc FeynRules}}\label{sec:fr}
\input fr.tex

\subsection{{\sc ALOHA}}\label{sec:aloha}
\input aloha.tex

\subsection{{\sc MadGraph~5}}
\input mg5.tex

\subsection{{\sc CalcHEP}}
\input calchep.tex

\section{\label{sec:pheno}Physics applications}

\subsection{Gravitino}\label{sec:gravitino}
\input gravitino.tex

\subsection{Top-quark excitation}\label{sec:top}
\input top.tex

\subsection{Angular distributions for a spin-$\mathbf{\tha}$ particle\label{sec:angular distributions}}
\input wigner.tex

\section{Summary}\label{sec:summary}
\input summary.tex

\begin{acknowledgement}{\textit{Acknowledgements}}
We would like to thank Fabio Maltoni for discussions in the early stage
 of the project.
N.D.C. would like to thank Nicholas Setzer and Daniel Salmon for helpful discussions.
N.D.C. was supported in part by the LHC-TI under U.S. National Science
 Foundation, grant NSF-PHY-0705682, by PITT PACC, and by the
 U.S. Department of Energy under grant No. DE-FG02-95ER40896.
PdA, KM and BO are supported inpart by the Belgian Federal Science Policy Office through the Interuniversity
Attraction Pole P7/37, in part by the ``FWO-Vlaanderen" through theproject G.0114.10N, and in part by the Strategic Research Program ``High Energy
Physics" and the Research Council of the Vrije Universiteit Brussel. 
O.M. is a fellow of the Belgian American Education Foundation. His work is partially supported by the IISN
MadGraph convention 4.4511.10. B.F. has been partially supported by the
Theory-LHC-France initiative of the CNRS/IN2P3, by the French ANR 12 JS05 002 01 BATS@LHC
and by the MCnet FP7 Marie Curie Initial Training Network.
C.D. is supported by the ERC grant ``IterQCD''. C.G. is supported by the
 Graduiertenkolleg ``Particle Physics at the Energy Frontier of New
 Phenomena''. Y.T. is supported by the Korea Neutrino Research
 Center which is established by the National Research Foundation of
 Korea (NRF) grant funded by the Korea government (MSIP) (No. 2009-0083526).
\end{acknowledgement}

\appendix
\section{\label{sec:app:conventions}Conventions}
\input conventions.tex

\section{\label{sec:app:polarization}Review of spin}
\input polarization_app.tex

\section{\label{sec:app:helgrav}Helicity amplitudes for gravitino pair-produciton}
\input helgrav.tex

\bibliography{library}
\bibliographystyle{spphys}

\end{document}

%% file: intro.tex
The recent discovery of the Higgs boson \cite{Aad:2012gk,Chatrchyan:2012gu} has greatly reinforced our expectation to find physics beyond the Standard Model (BSM) at the Large Hadron Collider (LHC).  For the first time, we have discovered a particle in Nature which is intrinsically unstable
with respect to quantum corrections and either requires unnaturally extreme fine-tuning or stabilization from a new sector of physics which will emerge at scales we will soon probe.  Although we do not yet know which BSM
theory will be the correct description of Nature, our current models often predict the presence of a spin-$\tha$ particle.  For example, in models of supergravity \cite{Deser:1976eh,Freedman:1976xh,Freedman:1976py,Ferrara:1976um,Cremmer:1978hn,Cremmer:1978iv,Cremmer:1982en,Cremmer:1982wb,Witten:1982hu}, the
graviton is accompanied by a spin-$\tha$ gravitino superpartner.
As another example, in models of compositeness, the top-quark has an associated spin-$\tha$ resonance
\cite{Burges:1983zg,Kuhn:1984rj,Kuhn:1985mi,Moussallam:1989nm,Almeida:1995yp,Dicus:1998yc,Walsh:1999pb,Cakir:2007wn,Stirling:2011ya,Dicus:2012uh,Hassanain:2009at}.
On different footings, if a spin-$\tha$ particle were discovered at the LHC, until the full picture was complete, an effective operator approach would be appropriate for determining its properties.  Because there are many
models that Nature could choose for the description of a spin-$\tha$ particle, it is essential that we simulate and analyze the signatures of these particles from as wide a class of models as possible, in preparation for the day we find it at an experiment.

There is an assortment of well-established Monte-Carlo pac\-ka\-ges used for simulating the collisions of high energy
phy\-sics, each with its own strengths.  Among these are {\sc CalcHEP}
\cite{Pukhov:1999gg,Boos:2004kh,Pukhov:2004ca,Belyaev:2012qa}, {\sc
MadGraph} 
\cite{Stelzer:1994ta,Maltoni:2002qb,Alwall:2007st,Alwall:2008pm,Alwall:2011uj}, {\sc Sherpa} \cite{Gleisberg:2003xi,Gleisberg:2008ta}, and {\sc Whizard} \cite{Moretti:2001zz,Kilian:2007gr} at parton level which are often followed by
{\sc Pythia} \cite{Sjostrand:2006za,Sjostrand:2007gs}, {\sc Sherpa}, or
{\sc Herwig} \cite{Corcella:2000bw,Bahr:2008pv} for radiation and
hadronization. 
Historically, these packages only supported the Standard Model (SM) and
a very small subset of the models BSM.  Until recently, spin-$\tha$
particles were not supported at all \cite{Belyaev:2012qa,Kilian:2001qz,Hagiwara:2010pi}.
The difficulty was that each model had to be implemented directly into the code of the respective Monte-Carlo packages.  This process required: first, an intimate knowledge of the Monte-Carlo package code; second, the hand-coding of thousands of lines of code for the Feynman rules and the parameter's dependence; and third, a long and tedious process of debugging.  Fortunately, the status of implementing new models into these simulation packages has recently and dramatically improved.  The first of these improvements is that the parton-level matrix element generators have begun to establish more general model input formats that require less intimate knowledge of their internal code.  The second improvement is the presence of packages that allow the user to enter the Lagrangian of their model rather than the individual vertices.  These packages then calculate the Feynman rules from the Lagrangian and export the model to the supported Monte-Carlo package of choice.  The first of these packages was {\sc LanHEP} \cite{Semenov:1996es,Semenov:1998eb,Semenov:2002jw,Semenov:2008jy,Semenov:2010qt} followed more recently by {\sc FeynRules} \cite{Christensen:2008py,Christensen:2009jx,Christensen:2010wz,Duhr:2011se,Fuks:2012im,Alloul:2013fw} and {\sc SARAH} \cite{Staub:2008uz,Staub:2012pb}.  

Although the status for implementing new models into Monte-Carlo packages has greatly improved, the situation for spin-$\tha$ has still been lacking.  The purpose of this paper is to report the implementation of support for spin-$\tha$ particles in {\sc FeynRules}, in its {\sc CalcHEP} export interface, in its Universal {\sc FeynRules} Output ({\sc UFO})
export interface, in the {\sc UFO} format \cite{Degrande:2011ua}, in the Automatic Libraries Of Helicity Amplitude ({\sc ALOHA}) package \cite{deAquino:2011ub} and in {\sc MadGraph}~5.
We further use this chain of packages to implement the gravitino of
supergravity together with its interactions and a model inspired by
quark and lepton compositeness involving a spin-$\tha$ quark excitation.
These implementations are then employed for reproducing several physics
results of earlier studies. 

This paper is organized as follows:
Section~\ref{sec:spin 3/2 intro} gives a brief introduction to spin-$\tha$ fields.
Section~\ref{sec:implement} describes the implementation of spin-$\tha$ fields in
{\sc FeynRules}, {\sc UFO}, {\sc ALOHA}, {\sc MadGraph}~5 and {\sc CalcHEP}.
Section~\ref{sec:pheno} presents the phenomenological applications of these implementations for their validation.
Section~\ref{sec:summary} summarizes our results.
We collect our conventions on Pauli and Dirac matrices, epsilon tensors, {\it etc.}, in
Appendix~\ref{sec:app:conventions} and include a review of spin, together with a detailed discussion of field polarization vectors and propagators, in Appendix~\ref{sec:app:polarization}.
Helicity amplitudes associated with gravitino pair-production, relevant for the discussion of
Section~\ref{sec:gravitino_grvgrv}, are given in Appendix~\ref{sec:app:helgrav}.

\section{\label{sec:spin 3/2 intro}Spin-$\mathbf{\tha}$ fermion fields} 

The first relativistic description of spin-$\tha$ fermions was carried
out by Rarita and Schwinger in 1941 \cite{Rarita:1941mf}. This was achieved by introducing a
generalized version of the Dirac equation, in which the spin-$\tha$
particles were described by a Dirac spinor with a Lorentz index, the
so-called Rarita-Schwinger field. In modern notation, the free
classical equation of motion for this field reads
\begin{equation}
 (\epsilon^{\mu\nu\rho\sigma}\gamma_5 \gamma_\rho \partial_\sigma
 + 2iM \gamma^{\mu\nu})\Psi_\nu =0 \;,
\label{raritaschwingereq}
\end{equation}
where $M$ is the mass of the spin-$\tha$ fermion described by
$\Psi_\nu$, and the Lorentz generators in the (four-component) spinorial 
representation $\gamma^{\mu\nu}$ have been defined in Appendix~\ref{sec:app:conventions}.
In the language of group theory,  the Rarita-Schwinger
field belongs to the direct product of two representations of the
Lorentz group:  the vector representation $(\oh,\oh)$, which describes
objects with one Lorentz index, and the Dirac representation $(\oh,0)
\oplus (0,\oh)$, which describes a Dirac pair of spin-$\oh$
fermions. Altogether, this direct
product contains one Dirac pair of spin-$\tha$ fermions and two Dirac
pairs of spin-$\oh$ fermions,
\begin{eqnarray}
\scriptsize
&\left(\frac{1}{2},\oh \right) \otimes \left[\left(\oh,0\right) \oplus
\left(0,\oh\right)\right] =\hspace{1.5in}& \\
&\hspace{1in}\left(\oh,1\right) \oplus \left(\oh,0\right) \oplus \left(1,\oh\right) \oplus \left(0,\oh\right)\ . &\nonumber
\label{decRar}
\end{eqnarray}
We observe that there is an ordinary Dirac spinor $(\oh,0)
\oplus (0,\oh)$ in this direct product. This piece can be identified
with $\gamma^\nu \Psi_\nu$, which transforms as a Dirac spinor,
and can thus be removed by requiring that $\gamma^\nu \Psi_\nu = 0$. 
Furthermore, the $(\oh,1) \oplus (1,\oh)$ piece is reducible under the
little group and contains a pair of spin-$\oh$ fermions in addition to
the pair of spin-$\tha$ fermions that we are after.  This last pair of
spin-$\oh$ fermions can be identified with $\partial^\nu\Psi_\nu$,
which again transforms as a Dirac spinor.  This piece can be
removed by requiring $\partial^\nu\Psi_\nu=0$.  
Both of these equations, as well as the Dirac equation, are a result
of the free classical equation of motion.  That is,
Eq.~(\ref{raritaschwingereq}) implies
\begin{equation}\label{eq:reduce to 32}
 \gamma^\nu\Psi_\nu=0\ , \hspace{20pt} \partial^\nu \Psi_\nu = 0\ ,
 \hspace{20pt}  (-i \slashed{\partial}+M) \Psi_\nu = 0\ .
\end{equation}
Consequently, on-shell, the free classical equations of motion remove
the spin-$\oh$ fermions so that only the spin-$\tha$ fermions are left.  
However, since classical equations of motion, as well as the
resulting Eq.~(\ref{eq:reduce to 32}), only apply on-shell,
the two pairs of spin-$\oh$ fermions can be present in the case of
off-shell spin-$\tha$ fields.

An important example of a (Majorana)
spin-$\tha$ field is in supergravity theories where supersymmetry requires the spin-$2$ graviton to have a superpartner, the gravitino, with spin $\tha$.  
In such models, it is often more convenient to work in a two-component notation instead of the four-component notation introduced by Rarita and Schwinger.
In this case, the free field Lagrangian reads
\be\bsp
  {\cal L} =& \frac12 \epsilon^{\mu\nu\rho\sigma} 
    (\psi_\mu \sigma_\rho \del_\sigma\psibar_\nu +
     \del_\sigma \psi_\nu \sigma_\rho \psibar_\mu) \\ 
   & + i M (\psi_\mu\sigma^{\mu\nu}\psi_\nu + \bar\psi_\mu\bar\sigma^{\mu\nu}\bar\psi_\nu) \ ,
\esp\ee
where, now, $\psi_\nu$ ($\bar\psi_\nu$) represents the left-chiral (right-chiral) two-component
spin-$\tha$ field. We recall that our conventions of the generators of the Lorentz
algebra in the (two-component) spinorial representation are summarized in Appendix~\ref{sec:app:conventions}.
This Lagrangian leads
to the two-component form of Eq.~\eqref{eq:reduce to 32}, after replacing the Dirac matrices
by the Pauli matrices and the Dirac equation by the Weyl equation.

%% file: fr.tex

{\sc FeynRules} \cite{Christensen:2008py,%
Christensen:2009jx,Christensen:2010wz,Duhr:2011se,Fuks:2012im,Alloul:2013fw}
is a {\sc Mathematica} package that allows the user to 
compute Feynman rules for a model based on quantum field theory directly from the 
associated Lagrangian. It can be used for any model that satisfies 
locality as well as Lorentz and gauge invariance. More precisely, a 
necessary condition for {\sc FeynRules} to work correctly is that all 
indices appearing in the Lagrangian are contracted. Moreover, no 
assumption is made on
the dimensionality of the operators that appear inside the Lagrangian. This 
feature is particularly useful in the present case, as Lagrangians for spin-$\tha$
 particles necessarily involve operators of dimension greater than four.

The output of {\sc FeynRules} is generic enough to be exported into any 
other model format. Translation interfaces hence exist to convert the 
Feynman rules computed by {\sc FeynRules} into a format readable by 
various Feynman diagram generators. Currently such translation interfaces 
have been developed for {\sc Calc\-HEP} and {\sc Comp\-HEP}, {\sc Feyn\-Arts} and {\sc Form\-Calc} 
\cite{Hahn:1998yk,Hahn:2000kx,Hahn:2009bf,Agrawal:2011tm}, 
{\sc Mad\-Graph}~4, {\sc Sher\-pa} as well as 
{\sc Whi\-zard} and {\sc O'\-Me\-ga}. 
In addition, the user has the 
possibility to output the Feynman rules in the so-called Universal {\sc 
FeynRules} Output (UFO), a format for the implementation of new physics 
models into Feynman diagram generators that is not tied to any specific 
matrix element generator \cite{Degrande:2011ua}.

In the rest of this section, we describe how particles of spin-$\tha$ can be 
implemented into {\sc FeynRules} both in non-su\-per\-sym\-me\-tric theories as well as in
su\-per\-sym\-me\-tric theories.
In the following, we also detail the extension of the {\sc UFO} (Section~\ref{sec:ufo}) and {\sc CalcHEP}
(Section~\ref{sec:calchep}) interfaces, respectively.

The implementation of spin-$\tha$ particles into {\sc FeynRules} models follows
the same logic as the implementation of all other particles. We
therefore refer to the {\sc FeynRules} manual \cite{Christensen:2008py} for 
more details on the declaration of particle classes. 
One important difference with the manual is that it is now possible to implement both two- and 
four-component fermions \cite{Duhr:2011se}. Four-component spin-$\tha$ fields correspond to
the newly introduced particle class {\tt R}, while two-component fields are 
implemented by declaring instances of the class {\tt RW}. All the attributes
common to the other particle classes are available for spin-$\tha$ fields, 
together with the option {\tt Chirality} for two-component fields so that 
one can define both left and right-handed two-component spin-$\tha$ 
fields. Furthermore, each spin-$\tha$ field always carries by default an 
index of type {\tt Spin} (respectively {\tt Spin1} and {\tt Spin2} for 
left-handed and right-handed two-component fields) as well as an index of 
type {\tt Lorentz}. We stress that, when writing down a Lagrangian involving
spin-$\tha$ fields, it is mandatory for the {\tt Spin} and {\tt Lorentz} 
indices to be the first two indices (in this order) carried by the spin-$\tha$ 
fields.

In supersymmetric theories, 
spontaneous supersymmetry breaking predicts the existence of a massless
fermion, dubbed the goldstino. The way this field interacts 
with the rest of the particle spectrum can be derived from the conservation 
of the supercurrent. The latter is defined by the variation of
the model Lagrangian ${\cal L}$ under a supersymmetry transformation with
a (Majorana fermionic) parameter $(\e,\ebar)$, 
\be
  \delta_\e {\cal L} = \del_\mu K^\mu = \del_\mu \bigg[
     {\del {\cal L} \over \del \big(\del_\mu X\big)} \delta_\e X 
     \bigg] \ .
\ee
We have used Euler-Lagrange equations to simplify the right-hand side of
this equation, and
a sum 
over all fields $X$ of the theory is implicit.
Consequently, the 
(Majorana fermionic) supercurrent $(J^\mu,\Jbar^\mu)$ is defined by
\be
  \e\!\cdot\!J^\mu + \ebar\!\cdot\!\Jbar^\mu =  {\del {\cal L} \over \del
\big(\del_\mu X\big)} \delta_\e X - K^\mu \ .
\ee
This is a conserved quantity since $\partial_\mu J^\mu=0$ and $\partial_\mu\Jbar^\mu=0$.
Following standard superspace techniques, $J^\mu$ can be computed as 
\cite{Wess:1978ns,Iliopoulos:1974zv,FuksRausch}
\be\bsp
  J^\mu_\alpha =&\
      \sqrt{2} D_\nu\phi^\dag_i \big(\sigma^\nu
         \sibar^\mu  \psi^i \big)_\alpha   
    + \sqrt{2}i \big(\sigma^\mu  \psibar_i\big)_\alpha W^{\star i}
\\ &\
    + g \phi^\dag T_a \phi \big(\sigma^\mu\lambar^a\big)_\alpha
    - \frac{i}{2} \big(\sigma^\rho \sibar^\nu \sigma^\mu \lambar_a\big)_\alpha
        v^a_{\nu\rho} \ ,
\esp \ee
where sums over the model gaugino fields $\lambda$, two-com\-po\-nent 
fermions
$\psi$, scalar fields $\phi$ and vector bosons $v_{\mu}$ (the associated field 
strength tensors being denoted by $v_{\mu\nu}$) are implicitly accounted for.
Moreover, we generically denote by $D_\mu$ gauge covariant derivatives, 
by $T^a$ and $g$ representation matrices and the coupling constant of the gauge group and by $W^{\star i}$ 
the derivatives of the hermitian conjugate superpotential 
$W^{\star i}\equiv \partial W^\star(\phi^\dag) / \partial \phi^\dag_i$.

The supercurrent can be calculated in {\sc FeynRules} by 
issuing, in a {\sc Mathematica} session,
\begin{verbatim}
  SuperCurrent[lv,lc,lw,sp,mu]
\end{verbatim}
where the variables \texttt{lv}, \texttt{lc} and \texttt{lw} are connected
to the different pieces of the supersymmetric Lagrangian ${\cal L}_v$, ${\cal L}_c$
and ${\cal L}_w$ defined below, and {\tt sp}
and {\tt mu} are spin and Lorentz indices, respectively. 
The three Lagrangians above must be given as full series expansions in terms of the Grassmann variables and read
\be\bsp
  {\cal L}_v = &\  \frac{1}{16 g^2} \Big[ W^{a\alpha} W_{a\alpha} +
      \bar W^a_\alphadot \bar W_a^\alphadot \Big] \ , \\
  {\cal L}_c = &\ \Phi^\dag e^{-2gV} \Phi \ , \\
  {\cal L}_w = &\ W(\Phi) \ , 
\esp\ee
where sums over the whole chiral ($\Phi$) and vector ($V$) superfield 
content of the model are again understood. In these expressions, we have also 
introduced the superpotential $W(\Phi)$ and the superfield strength
tensors $W_\alpha$ and $\bar W_\alphadot$.  These three superfields
can be automatically obtained in {\sc FeynRules} by issuing
\begin{verbatim}
 lc = GrassmannExpand[ CSFKineticTerms[] ]; 
 lv = GrassmannExpand[ VSFKineticTerms[] ];
 lw = GrassmannExpand[ SPot+HC[SPot] ];
\end{verbatim}
where the superpotential {\tt SPot} is the only quantity to be provided by
the user, in addition to all the model (super)field declarations. 
We refer to Ref.\ \cite{Duhr:2011se} for more information.

Once extracted, the supercurrent can be further employed in the
construction of effective Lagrangians for the goldstino and gravitino fields
\cite{Fayet:1977vd,Fayet:1979yb}.
In this context, one or several of the
auxiliary fields of the supersymmetric model under consideration
acquire vacuum expectation values.
Supercurrent conservation is then understood  as the equations of motion for
the goldstino field. More explicitly, we take the example of a chiral
supermultiplet $(\phi_1,\psi_1,F_1)$ whose auxiliary field
$F_1$ gets a vacuum expectation value
$v_1/\sqrt{2}$. 
Supercurrent conservation then leads to
\be
  0 = \partial_\mu J^\mu = i v_1 (\sigma^\mu\partial_\mu \bar\psi_1)_\alpha
  + \partial_\mu \tilde J^\mu_\alpha \ ,
\label{eq:supercons}\ee
after having shifted the field $F_1$ by its vacuum expectation value,
the quantity $\tilde J$ containing all the other contributions to the supercurrent.
Considering Eq.~\eqref{eq:supercons}
as the equations of motion for the goldstino field identified here as $\psi_1$,
one deduces the corresponding Lagrangian
\be\bsp
  {\cal L}_{\psi_1} =&\ \frac{i}{2} \Big(\psi_1\sigma^\mu\partial_\mu\bar\psi_1 - 
   \partial_\mu\psi_1 \sigma^\mu\bar\psi_1\Big)
\\ &\  
  + \frac{1}{2 v_1} \Big[ \psi_1 \!\cdot\! \partial_\mu\tilde J^\mu 
  + \psibar_1 \!\cdot\! \partial_\mu\bar{\tilde J}^\mu \Big] \ .
\esp \ee

Once the goldstino has been eaten by the gravitino, the latter becomes massive and  the Lagrangian can be 
rewritten in terms of the (two-component) spin-$\tha$ gravitino field $\psi_G$,
\be\bsp
  {\cal L}_G =&\ \frac12 \epsilon^{\mu\nu\rho\sigma} 
    (\psi_{G\mu} \sigma_\rho \del_\sigma\psibar_{G\nu} +
     \del_\sigma \psi_{G\nu} \sigma_\rho \psibar_{G\mu})
\\ &\  
  + {\cal N} \Big[ \psi_{G\mu} \!\cdot\! J^\mu 
  + \psibar_{G\mu} \!\cdot\! \bar J^\mu \Big] \ . 
\esp \ee
where we have replaced the vacuum expectation value $v_1$ by the generic
normalization constant ${\cal N}$. In general, the latter is expressed in terms of
the su\-per\-sym\-me\-try-breaking scale and the gravitino mass.
Both Lagrangians derived above can subsequently be
implemented in {\sc FeynRules} by making use of standard
techniques.

\subsection{\sc{UFO}}\label{sec:ufo}

The {\sc{UFO}} format \cite{Degrande:2011ua} is, by construction, completely agnostic of the model and 
the associated interactions. In particular, the format does not make any
\textit{a priori} assumptions about the color and/or Lorentz structures that 
appear inside the vertices. Therefore, higher-dimensional operators can be 
implemented in the same way as renormalizable operators. 

On the other hand, the {\sc UFO} format was implicitly assuming a particular convention for propagators.
As explained in Appendix~\ref{sec:propagator}, the propagator for spin-$\tha$ is not unique and more than one expression are
commonly used in the literature. The same is true for the massless
propagator which depends on the gauge fixing term
\cite{VanNieuwenhuizen:1981ae}. We therefore extend the UFO format in order to support model specific propagators. 

Starting from the original UFO format, we include one additional file {\tt propagators.py} that contains definitions for
propagators. These should be instances of the {\tt Propagator} class which follows the usual conventions
as for other {\sc UFO} objects. It has two mandatory and one optional arguments:
\begin{itemize}
\item {\tt name} [mandatory]:  a short tag for identifying the propagator expression;
\item {\tt numerator} [mandatory]: an analytical expression of the numerator of the propagator provided
following the conventions of the {\sc UFO/ALOHA} format for Dirac matrices, four-momenta, {\it etc.};
\item {\tt denominator} [optional]: an analytical expression of the denominator of the propagator
  provided in the conventions of the {\sc UFO/ALOHA} format for Dirac matrices, four-mo\-men\-ta, {\it etc}.
If this is not provided, the Feynman propagator denominator $(p^2-M^2+iM\Gamma)$ is employed for
a particle of mass $M$ and with a four-momentum $p^\mu$.
\end{itemize}

The non-contracted indices of the numerator should be ``1" and ``2" (and ``51" and ``52" for spin-2).
In the presence of a flow, the index ``1" of the analytical expression  will be contracted with outgoing particles, while the index 2 will be contracted with incoming particles. If an expression requires the index of a particle, 
\textit{e.g.}, the momentum, the index should be {\tt id}.
For example, the propagators associated with a massless vector field and with a massless spin-$\tha$
fields,
\be\bsp
  \Delta^{1,\mu\nu}(p) = &\ \frac{1}{p^2+i\epsilon} \Big[-\eta^{\mu\nu} \Big] \ , \\
  \Delta^{3/2,\mu\nu}(p) = &\ \frac{1}{p^2+i \epsilon} \Big[- \gamma^\mu\slashed{p}\gamma^\nu \Big] \ ,
\esp\label{eq:propmassless}\ee
can be implemented as\footnote{For the sake of the example, we have included
the definition of the denominator in the \texttt{V0} implementation. This is however
not necessary as this consists of the standard Feynman propagator for a massless particle.}
\begin{verbatim}
V0 = Propagator(name = "V0", 
   numerator = "-1 * Metric(1, 2)", 
   denominator = "P(-1, id)**2")

R0 = Propagator(name = "R0", 
   numerator = "-1 * Gamma(1, 1,-1) * \
   PSlash(-1,-2, id) * Gamma(2,-2, 2)")           
\end{verbatim}

Furthermore, the {\tt Particle} class has been supplemented by a new optional attribute which
refers to the {\tt propagator} to associate with each instance of the class. Hence, the propagator
to employ can be referred to by its name as implemented in the file {\tt propagators.py}.
For example, a massless spin-$\tha$ particle with a custom propagator as given in
Eq.~\eqref{eq:propmassless} would be defined by\footnote{This 
assumes that, at the beginning of the file, the line ``{\tt import propagators}'' is present.}
\begin{verbatim}
grv0 = Particle(pdg_code = 1000039,
    name = 'grv0',
    antiname = 'grv0',
    spin = 4,
    color = 1,
    mass = Param.ZERO,
    width = Param.ZERO,
    propagator = propagators.R0
    texname = 'grv',
    antitexname = 'grv',         
    charge = 0,
    GhostNumber = 0,
    Y = 0)
\end{verbatim}
If the attribute {\tt propagator} is not present, the default propagator is employed.
In the case of massless spin-$\tha$ fields, it is given as in the second line of Eq.~\eqref{eq:propmassless}.
In the massive case, we refer to Eq.~\eqref{eq:sp32prop} for the propagator numerator
while the denominator consists of Feynman propagator denominator.

%% file: aloha.tex

{\sc ALOHA} \cite{deAquino:2011ub} is a {\sc Python} module which automatically computes Helicity Amplitude
Subroutines ({\sc Helas}) \cite{HELAS} associated with the interactions of a given model from its implementation in the UFO format.
These routines allow for a fast numerical computation of squared matrix
elements using helicity amplitudes, with which the complexity of
numerical computations grows only linearly with the number of diagrams.
This contrasts with trace techniques based on completeness relations
applied to squared amplitudes, where the complexity grows quadratically
with the number of diagrams. 
Support for spin-$\tha$ particles requires the implementation of the external wave functions \cite{Hagiwara:2010pi},
as well as
additional internal objects such as the spinorial representation and  default propagators.
In addition, the {\sc ALOHA} code has been made faster and more
flexible, and new features have been introduced to support for:
\begin{itemize}
\item a user-defined propagator [required for spin-$\tha$];
\item including model parameters in the analytical expressions of the helicity routines
  [required for spin-$\tha$];
\item conjugate routines \cite{Denner:1992vza} with different incoming and outgoing spins
  [required for spin-$\tha$];
\item division operation in the analytical expressions of the helicity routines;
\item mathematical functions (exponential, logarithm, \textit{etc.}) that have a single
(scalar) argument.
\end{itemize}

Furthermore, the {\sc ALOHA} outputs, {\it i.e.} the helicity routines, are better optimized. 
Firstly, all scalar invariants are identified and computed only once. Secondly, the 
final expressions are written in a way minimizing the number of computing operations required by the code.
These optimizations allow for a gain of a factor of two in speed, compared to the previous version of the
package.

The new {\sc ALOHA} code has been extensively
tested. First, results have been compared to the previous version of the code for large classes of models such
as the Standard Model, the Minimal Supersymmetric Standard Model, Randall-Sundrum extra-dimensional models,
effective theories with dimension-six operators, \textit{etc}. In each of these models, more than 6000
processes have been considered and checked. 
Additionally, we have compared the results obtained with
{\sc ALOHA} to known analytical formulas for the most complex cases, such as for effective vertices
associated with loop-induced interactions (\eg, the $Hgg$ coupling).

%% file: mg5.tex

The Monte Carlo package {\sc MadGraph}~5~\cite{Alwall:2011uj} is a suite of programs related to matrix element studies \cite{Maltoni:2002qb,Alwall:2008pm,Artoisenet:2012st,Artoisenet:2010cn,Hirschi:2011pa,Frederix:2009yq}.
Its most commonly used tools allow for generating the Feynman diagrams associated with
a specific scattering process, computing the related matrix element and cross section in a
very efficient way and eventually
generating events both at the leading order~\cite{Alwall:2011uj} and next-to-leading
order~\cite{Alwall:2013} in perturbation theory.
In this framework, the task left to the user consists mainly of  specifying:
\begin{itemize}
  \item the particle physics model under consideration, which can be either
    renormalizable or not, without any restriction on the dimensionality of the included operators;
  \item the process under consideration in
    terms of initial and final state particles;
  \item the collision basic setup which includes information such as
    the energy or the nature of the colliding beams;
  \item a set of kinematical selection criteria on the final state particles to be produced.
\end{itemize}

The program has been designed to be extremely flexible and generic with respect to the physics model
under consideration. To this aim, it relies both on the {\sc UFO} for its model format and on the {\sc ALOHA}
package. The latter allows to generate all helicity routines \cite{HELAS} necessary for the evaluation of the
amplitude associated with the considered process. We benefit from this flexibility and only
modify the program in a minimal way in order to implement spin-$\tha$ support. More into details,
two minor modifications of the code have been required. First, {\sc MadGraph}~5 has been upgraded to be capable
to deal with arbitrary propagators (see Section \ref{sec:ufo}). Next, information on the number of
degrees of freedom associated with a spin-$\tha$ particle is now included in {\sc MadGraph}~5.

The {\sc FeynRules} interface converting a model into a UFO library and the use of the latter within the 
{\sc MadGraph}~5 framework have been widely tested and validated in the context of several physics applications
in Section~\ref{sec:pheno}.

%% file: calchep.tex
\label{sec:calchep}

{\sc CalcHEP} \cite{Pukhov:1999gg,Belyaev:2012qa} is a software package which is designed for effective evaluation and simulation of high-energy physics collider processes.
The main features of {\sc CalcHEP} are the computation of Feynman diagrams, integration over multi-particle phase space and event simulation at parton level. The principle attractive key-points along these lines are that it has:
\begin{itemize}
  \item an easy startup even for those who are not familiar with {\sc CalcHEP}; 
  \item a friendly and convenient graphical user interface (GUI); 
  \item the option for a user to easily modify a model or introduce a new model by either using the graphical interface or by using an external package with the possibility of cross checking the results in different gauge choices; 
  \item a batch interface which allows to perform very complicated and tedious calculations connecting production and decay modes for processes with many particles in the final state.
\end{itemize}
With this features set, {\sc CalcHEP} can efficiently perform calculations with a high level of automation from a theory in the form of a Lagrangian down to phenomenology in the form of cross sections, parton level event simulation and various kinematical distributions.

{\sc CalcHEP} was written to accept very general operators since its
creation.  Including spin-$\tha$ support mainly involved the inclusion
of the Rarita-Schwinger propagator\footnote{The massive spin-$\tha$ and spin-$2$ propagators in {\sc CalcHEP} have the form given in Appendix~\ref{sec:propagator}.  However, the propagator formulas given in Sec.~8.6 of Ref.~\cite{Belyaev:2012qa} must
be divided by $3$ and $6$, respectively.  These factors were incorrectly entered in Ref.~\cite{Belyaev:2012qa}.}
 and allowing the {\sc CalcHEP} vertex to include
a Lorentz index for the spin-$\tha$ fermions \cite{Belyaev:2012qa}.
{\sc CalcHEP} calculates all its squared matrix elements symbolically.
At very low multiplicity, this allows {\sc CalcHEP} to maximally optimize the numerical code.  At very high multiplicity, the time and memory required to do this optimization can however become prohibitive.

The interface between {\sc FeynRules} and {\sc CalcHEP} \cite{Christensen:2009jx} was updated to support spin-$\tha$ particles as part of the current project.  The normalization and angular distributions computed by {\sc CalcHEP} were tested and found to agree with theories in this paper (see Section~\ref{sec:pheno}).

%% file: gravitino.tex
 
The most popular candidate for a spin-$\tha$ particle among new physics
scenarios might be the gravitino, the Majorana spin-$\tha$
superpartner of a graviton in local
supersymmetric extensions to the Standard Model \cite{Deser:1976eh,Freedman:1976xh,Freedman:1976py,Ferrara:1976um,Cremmer:1978hn,Cremmer:1978iv,Cremmer:1982en,Cremmer:1982wb,Witten:1982hu}. 
As briefly discussed in Section~\ref{sec:fr}, the
gravitino becomes massive via the super-Higgs mechanism by absorbing
the massless spin-$\oh$ goldstino. 
One of the key features of the gravitino is its mass $m_{3/2}$, 
which is related to the scale of supersymmetry breaking $M_{\rm SUSY}$ as well as to the Planck
scale $M_{\rm Pl}$,
\be
m_{3/2}\sim (M_{\rm SUSY})^2/M_{\rm Pl}\ . 
\ee
While the interactions of the gravitino/goldstino and the
associated {\sc Helas} subroutines~\cite{Hagiwara:1990dw}
were introduced manually for {\sc MadGraph}~4
in Refs.~\cite{Hagiwara:2010pi,Mawatari:2011jy} and tested carefully in
further phenomenological studies~\cite{Mawatari:2011cu,deAquino:2012ru}, 
now they can be automatically generated using the {\sc FeynRules} and {\sc
Aloha} packages (see
Section~\ref{sec:fr} and \ref{sec:aloha}) for generic event generators.

In this section, as a non-trivial test for our implementation, we study
the two interesting physics applications: the high-energy tree
unitarity violation in gravitino pair-production and the
gravitino contribution to gluino pair-production.

\subsubsection{Gravitino pair-production}
\label{sec:gravitino_grvgrv}

\begin{figure}
\center
 \includegraphics[width=0.45\columnwidth]{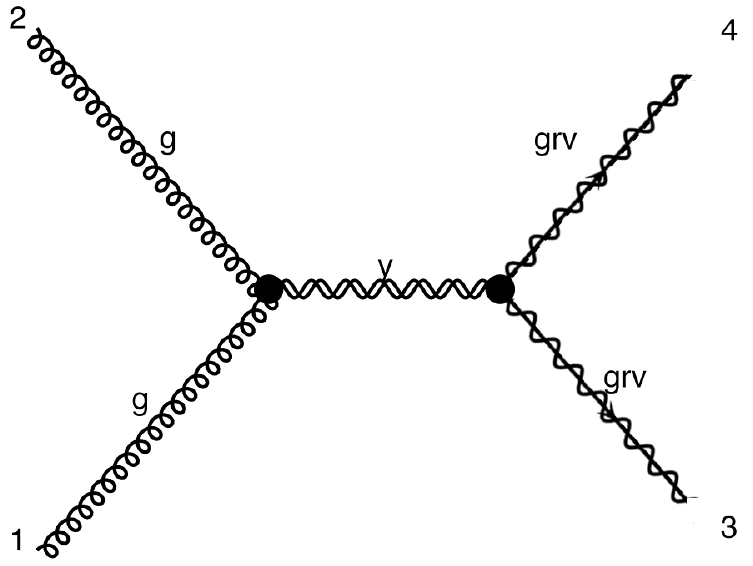}\quad
 \includegraphics[width=0.45\columnwidth]{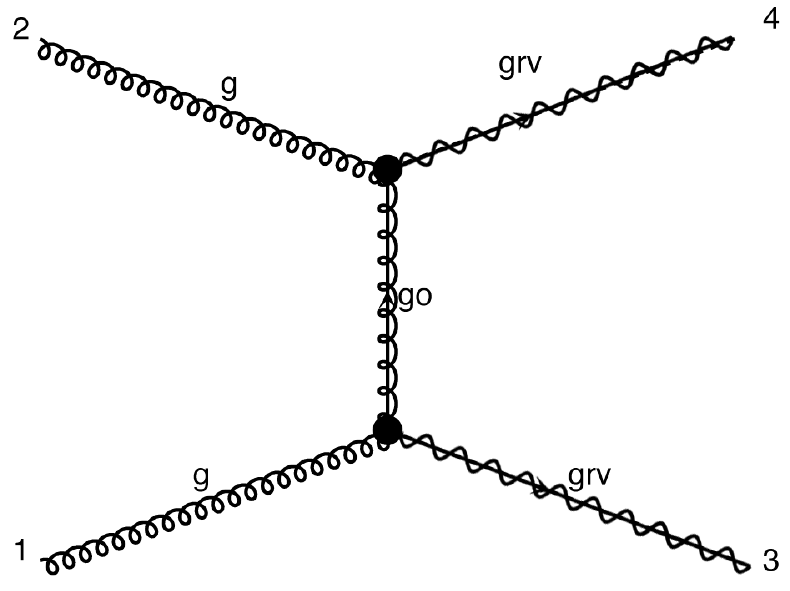}\\
 (a)\hspace*{4cm}(b)
\caption{The relevant diagrams contributing to the fastest energy growth
 in the $gg\to\gld\gld$ process. They consist of
 an $s$-channel graviton exchange diagram (a) and
 $t,u$-channel gluino exchange diagrams (b).}
\label{fig:diagrams}
\end{figure} 

Tree unitarity violation in the gravitino pair production process $gg\to\gld\gld$ has been studied by
Bhattacharya and Roy in Ref.~\cite{Bhattacharya:1988zp}.  In this section, we compare our
numerical results with their approximated analytic results.
There are three diagrams which contribute to the fastest energy growth
in this process, an
$s$-channel graviton exchange diagram and $t,u$-channel gluino exchange
diagrams, as shown in Figure~\ref{fig:diagrams}%
\footnote{There are also $s$-channel sgoldstino (the scalar
superpartner of the goldstino) exchange diagrams, which give rise to the
next-to-fastest energy growth.}.

The relevant interaction Lagrangian is given by
\be
 {\cal L}=
 -\frac{1}{2\Mpl}\overline{\Psi}_{\mu}\gamma^{\nu\rho}
  \gamma^{\mu}\lam_a g^{a}_{\nu\rho} 
 -\frac{1}{\Mpl}(T^{\mu\nu}_g+T^{\mu\nu}_{\gld})G_{\mu\nu},
\label{lag_grv}
\ee
where $\Psi_{\mu}$ is the spin-$\tha$ gravitino field, $\lambda_a$ the gluino field,
and $g^a_{\mu\nu}$ the field strength tensor for a gluon $g^a_{\mu}$. We also recall
that our conventions on the generators of the Lorentz algebra in the spinorial representation
$\gamma^{\mu\nu}$ are defined in Appendix~\ref{sec:app:conventions}. Moreover,
$\Mpl\equiv M_{\rm Pl}/\sqrt{8\pi}\sim 2.4\times 10^{18}$~GeV denotes the
reduced Planck mass.
The first term in the Lagrangian of Eq.~\eqref{lag_grv} has been extracted from the complete
set of gravitino interactions as described in
Section~\ref{sec:fr}, whereas its second
term describes the interactions of a spin-2 graviton $G_{\mu\nu}$ with 
gluons \cite{deAquino:2011ix} and gravitinos. It shows the well-known feature
that a spin-2 graviton couples to other fields via their
energy-momentum tensor,  $T_g^{\mu\nu}$ and $T_{\gld}^{\mu\nu}$ in the gluon and gravitino
cases, respectively. Those two last quantities are given by
\be\bsp
 T_{g}^{\mu\nu}=
  &\   \frac{1}{4}\eta^{\mu\nu}g_a^{\rho\sigma}g^a_{\rho\sigma}
      -g^{a,\mu}_\rho g_a^{\nu\rho} \\
  &\   -\frac{1}{\xi}\eta^{\mu\nu}\big\{g^a_\rho\,\partial^\rho\partial^\sigma
 g_{a,\sigma}
      +\frac{1}{2}(\partial^\rho g^a_\rho) (\partial^\sigma g_{a,\sigma})\big\}\\
  &\  +\frac{1}{\xi}\left( g^{a,\mu}\partial^\nu\partial^\rho g_{a,\rho} 
         +g^{a,\nu}\partial^\mu \partial^\rho g_{a,\rho} \right) \ , \\ 
 T^{\mu\nu}_{\gld}=&\
   \frac{1}{8}\epsilon^{\lambda\rho\sigma\mu}\overline{\Psi}_{\lambda}\gfive\gamma^{\nu}\overleftrightarrow{{\partial}_{\sigma}}\Psi_{\rho}+(\mu\leftrightarrow\nu)\\
 &\ -\frac{i}{8}\epsilon^{\lambda\rho\sigma\mu}\partial_{\tau}(\overline{\Psi}_{\lambda}\gfive\{\gamma_{\sigma},\gamma^{\nu\tau}\}\Psi_{\rho})
 +(\mu\leftrightarrow\nu)\\
 &\ -im_{3/2}\overline{\Psi}_{\lambda}(\eta^{\mu\nu}\gamma^{\lambda\rho}-\eta^{\rho\nu}\gamma^{\lambda\mu}-\eta^{\rho\mu}\gamma^{\lambda\nu})\Psi_{\rho} \ ,
\esp\ee
where the dependence on the gauge fixing parameter $\xi$ in the gluon energy-momentum tensor has been
kept explicit although we adopt Feynman gauge, \ie, $\xi=1$, in the rest of this section.
Moreover, the operator $\overleftrightarrow{\partial_\sigma}$ stands for the usual
Hermitian derivative operator,
\be
\overline{\Psi}_{\lambda}\cdots\overleftrightarrow{\partial}_{\sigma}\Psi_{\rho}
 = \overline{\Psi}_{\lambda}\cdots\partial_{\sigma}\Psi_{\rho} -\left(\partial_{\sigma}\overline{\Psi}_{\lambda}\right)\cdots\Psi_{\rho}\ .
\ee

To validate our gravitino implementation,
we compute explicitly the helicity amplitudes ${\cal M}_{\lam_1\lam_2,\lam_3\lam_4}$
associated with the process
\be
 g(p_1,\lam_1)+g(p_2,\lam_2)\to\gld(p_3,\lam_3)+\gld(p_4,\lam_4)\ .
\ee
They are presented in Appendix~\ref{sec:app:helgrav}.
We then select a particular helicity combination and choose to investigate
the properties of the ${\cal M}_{1,-1,\frac{1}{2},-\frac{1}{2}}$ amplitude,
as it gives rise to the leading energy growth with the partonic center-of-mass (CM)
energy $\sqrt{s}$. 
Expanding the $s$-, $t$- and $u$-channel contributions
to the considered amplitudes in terms of the gravitino mass, one gets
\be\bsp
 {\mathcal{M}}^s = &\ -\frac{s}{12\Mpl^2}(1+c_\theta)s_\theta \Big[ 
   \frac{s}{m^2_{3/2}} +4 \Big] + {\cal O}(m_{3/2}^2) \ ,\\
 {\mathcal{M}}^t = &\  \frac{s}{6\Mpl^2 x_t} s^3_\theta  + {\cal O}(m_{3/2}^2) \ ,\\
 {\mathcal{M}}^u = &\  \frac{s}{24\Mpl^2x_u}(1+c_\theta)^2s_\theta
    \Big[ \frac{s}{m^2_{3/2}} + 1 + \frac{1}{x_u}(1+c_\theta) \Big]\\
   &\quad  + {\cal O}(m_{3/2}^2) \ ,
\esp\ee
following the notations of Appendix~\ref{sec:app:helgrav}, 
and where $s_\theta$ and $c_\theta$ stand for the sine and cosine of the scattering angle $\theta$
defined as the angle between the $\mathbf p_1$ and $\mathbf p_3$
directions in the partonic center-of-mass frame. 
We also introduce the reduced variables
\mbox{$x_{t,u}=\frac{m_{\go}^2}{s}+\frac{1}{2}(1\mp c_\theta)$}.
In the limit of large center-of-mass energies, or equivalently when
$m_{\go}/\sqrt{s}\to0$, the $s^2$ energy growth behavior of the $s$- and $u$-channel
diagrams cancels,
\be\bsp
 {\mathcal{M}}_{1,-1,\frac{1}{2},-\frac{1}{2}} \to
  -\frac{s}{6\Mpl^2}s_\theta \Big[ 
     \frac{m_{\go}^2}{m_{3/2}^2} \!-\! \frac{3}{2}(1\!+\!c_\theta) \Big] 
  \!+\! {\cal O}(m_{3/2}^2) \ ,
\esp
\label{eq:approxM}
\ee
whereas the linear $s$-dependence still remains~\cite{Bhattacharya:1988zp}.

\begin{figure}
\center
 \includegraphics[width=0.9\columnwidth]{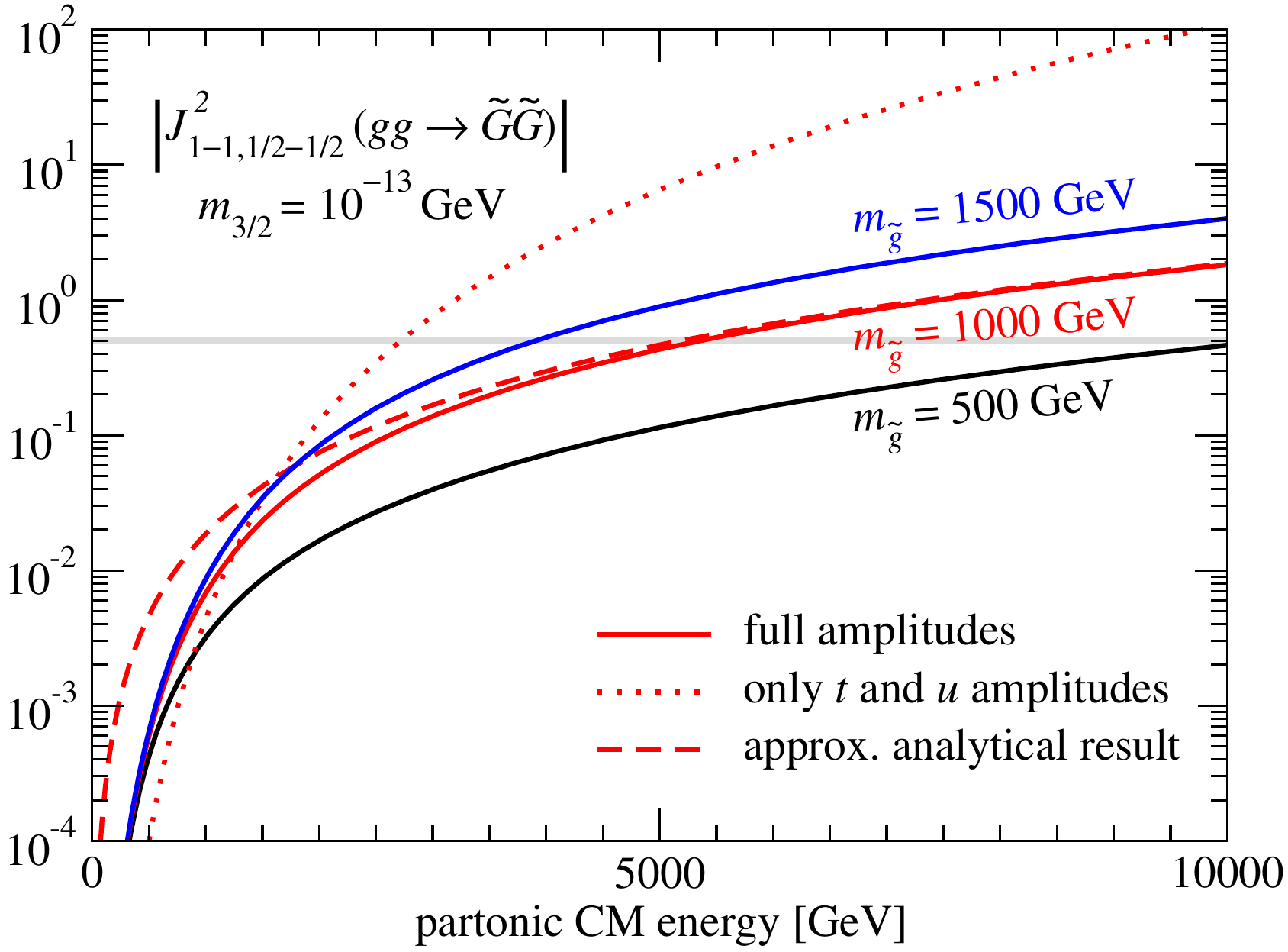}\\
\caption{Energy dependence of the projected partial wave amplitude 
 $\big|J^2_{1-1,\frac{1}{2}-\frac{1}{2}}\big|$ defined in Eq.~(\ref{pamp})
 for the $gg\to\gld\gld$ process and for different choices of the gluino mass.
 The solid lines are calculated by
 {\sc MadGraph}~5 and use the full amplitude. For the $m_{\go}=1000$~GeV case,
 we also show the analytical results in the
 $\sqrt{s}\gg m_{3/2},m_{\go}$ limit of Ref.~\cite{Bhattacharya:1988zp} (dashed)
  as well as the $s^2$ dependence resulting from the $t$- and
  $u$-channel diagram contributions only (dotted).}
\label{fig:sdep_godep}
\end{figure} 

One of the advantages of our implementation consists of flexibility,
\ie, one can vary parameters in the entire model parameter space and efficiently
produce the related results. Figure~\ref{fig:sdep_godep} hence illustrates
the energy dependence of the projected partial wave amplitude,
\be
 J^2_{1,-1,\frac{1}{2},-\frac{1}{2}}
 =\frac{1}{32\pi}\int_{-1}^{1} \d\cos\theta\, d^2_{21}(\theta)\,
  {\cal M}_{1,-1,\frac{1}{2},-\frac{1}{2}} \ ,
\label{pamp}
\ee
for different gluino masses.
In this expression, we introduce the $J=2$ Wigner $d$-function
$d^2_{21}(\theta)$ (see Ap\-pen\-dix~\ref{sec:Wigner d-functions}).
The analytic result of Eq.~(\ref{eq:approxM}) (dashed line)
is derived
in the $\sqrt{s}\gg m_{3/2},m_{\go}$ limit~\cite{Bhattacharya:1988zp} and agree with our numerical
predictions calculated by {\sc MadGraph 5} in the high-energy region
where this approximation is valid%
\footnote{{\sc CalcHEP} does not currently support massless spin-2 particles, so we were not able to compare either the
symbolic or numerical results with {\sc CalcHEP}.}.
%
In contrast, in the low-energy regime, the sub-leading term to
Eq.~(\ref{eq:approxM}) in the $m^2_{\tilde{g}}/s$ expansion becomes relevant, and one observes the
deviation of the approximated analytical result from our numerical predictions. 
 Finally, partial wave unitarity requires
\be
 \big|J^2_{1,-1,\frac{1}{2},-\frac{1}{2}}\big| < \frac{1}{2} \ ,
\ee
which is shown by a gray line on the figure.

We note that, in the $\sqrt{s}\gg m_{3/2}$ limit, goldstino-gravitino equivalence allows us to replace the
spin-$\tha$ gravitino by the spin-$\oh$ goldstino (which we
denote by $\chi$), \mbox{$\Psi_{\mu}\sim\sqrt{2/3}\,\del_{\mu}\chi/m_{3/2}$}.
We have verified that we obtain results similar to those of Figure~\ref{fig:sdep_godep} in this case.

\subsubsection{Gluino pair-production}

\begin{figure}
\center
 \includegraphics[width=0.9\columnwidth]{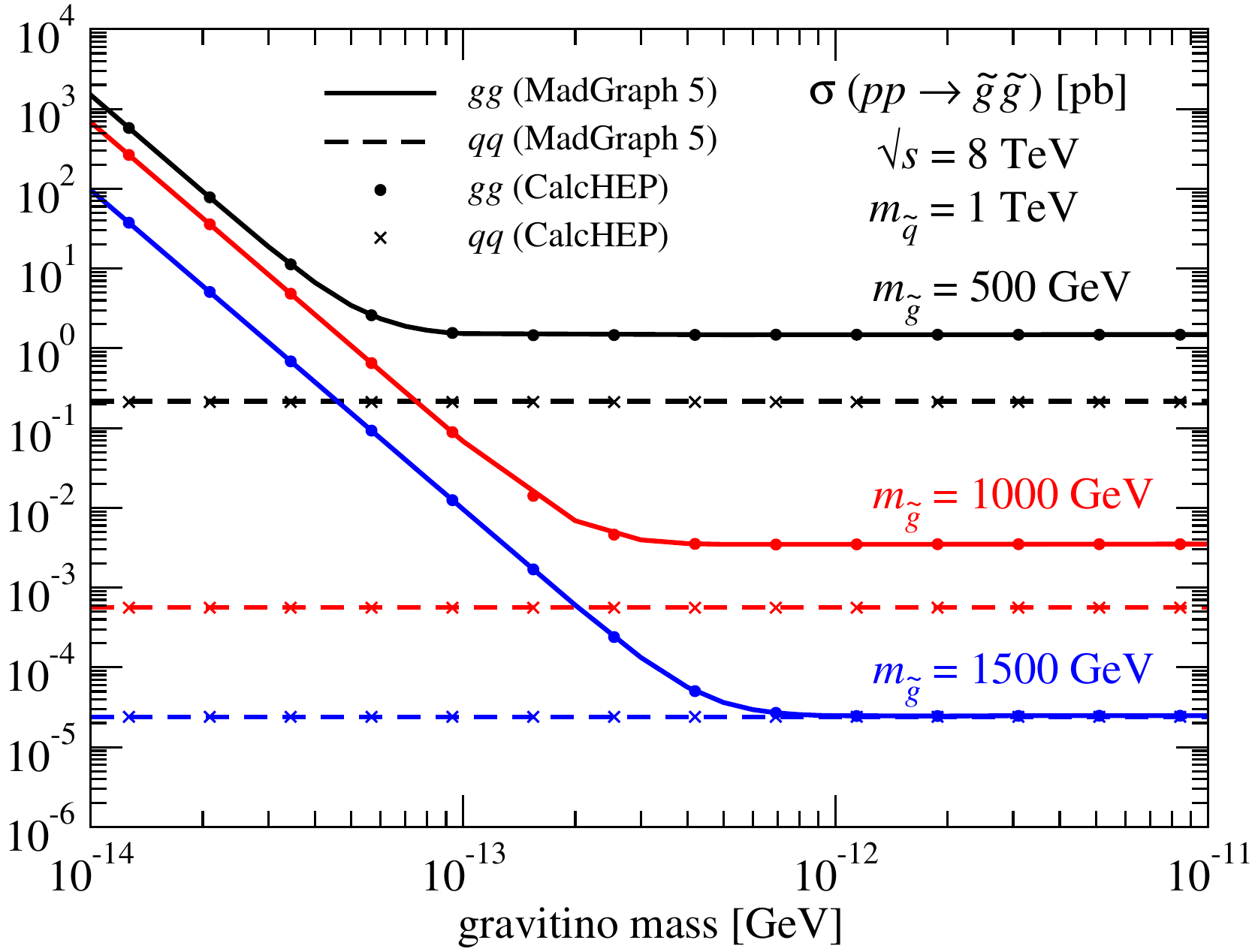}\\[2mm]
 \includegraphics[width=0.9\columnwidth]{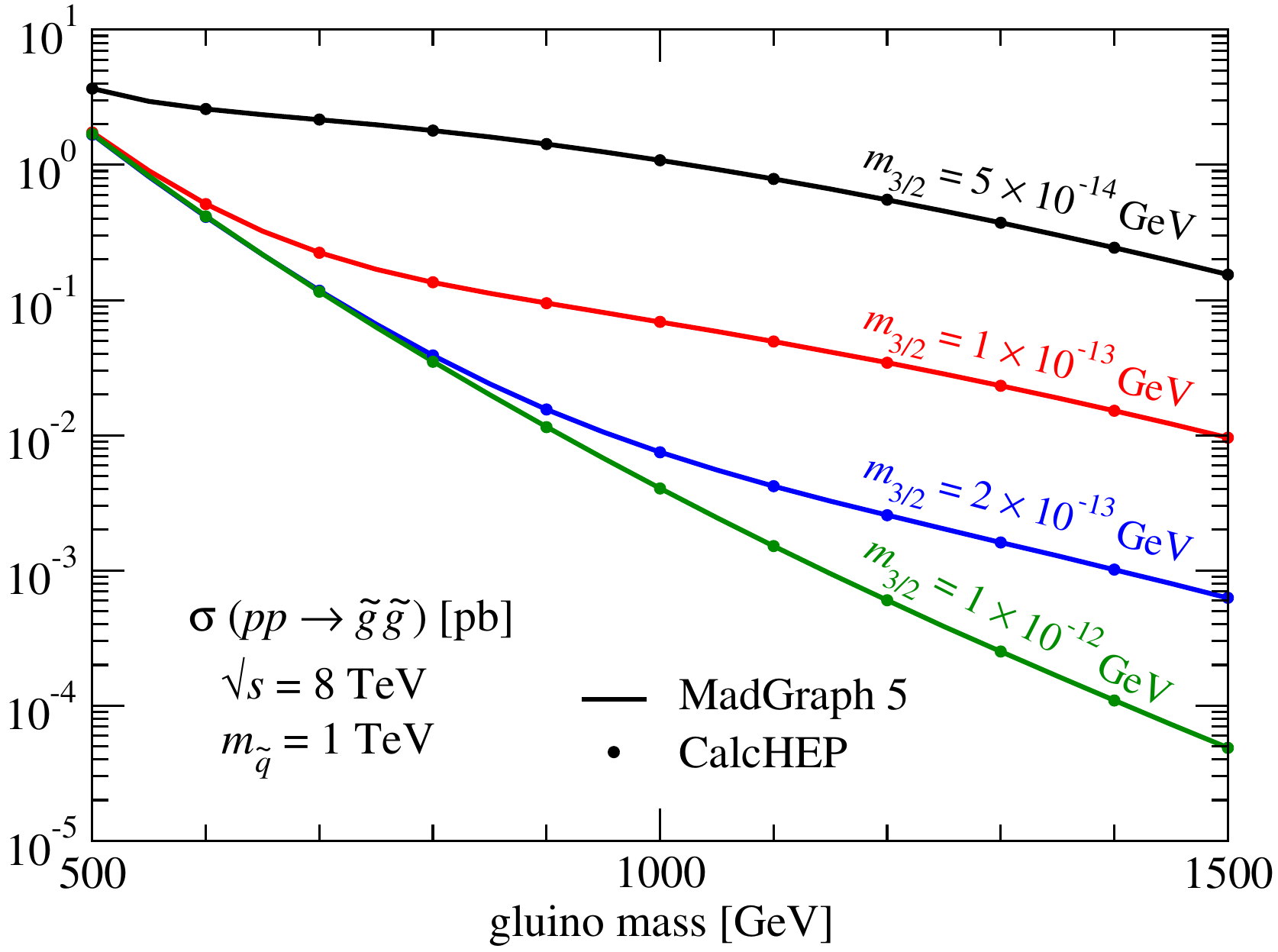}
\caption{Total cross sections for gluino pair-production at the LHC, running
  at a center-of-mass energy of 8~TeV, presented as a function of the gravitino mass (upper panel) and of
 the gluino mass (lower panel). Squark masses are fixed to 1~TeV.}
\label{fig:gluinopair}
\end{figure}

We now turn to the analysis of the $t,u$-channel gravitino 
contributions to gluino pair-production when the final state
arises from gluon scattering~\cite{Dicus:1989gg,Kim:1997iwa}.

In Figure~\ref{fig:gluinopair}, we present the different contributions
to the total cross section for gluino pair-production
$pp\to\go\go$ at the LHC, running at a center-of-mass energy of 8~TeV.
In the upper panel of the figure, we fix the squark masses to 1~TeV and investigate the dependence of the cross
section on the gravitino mass,
for several choices of the gluino mass.
The solid lines refer to the
contribution from gluon scattering while the dashed lines are related to
the quark-antiquark one.
The gravitino diagrams contribute significantly only when the gravitino is
very light with respect to the gluino mass. In other words, when $m_{3/2} \lesssim 6\cdot 10^{-14}$~GeV,
$3\cdot 10^{-13}$ GeV, and $7\cdot 10^{-13}$ GeV for a gluino of 500~GeV, 1000~GeV and
1500~GeV, respectively.

In the lower panel of Figure~\ref{fig:gluinopair}, we analyze the dependence of the total cross
section on the gluino mass for squark masses of 1~TeV and various values of the gravitino mass.
This illustrates the scaling of the cross section in
$1/m^4_{3/2}$ in the parameter space regions where gravitino diagrams dominate the cross section.

%% file: top.tex
Among the large number of new physics theories, those constructed
upon the idea of compositeness of the Standard Model quarks and leptons usually
predict the existence of both spin-$\oh$ and spin-$\tha$ states 
\cite{Burges:1983zg,Kuhn:1984rj,Kuhn:1985mi,Moussallam:1989nm,Almeida:1995yp,%
Dicus:1998yc,Walsh:1999pb,Cakir:2007wn,Stirling:2011ya,%
Dicus:2012uh,Hassanain:2009at}.
Since the top quark, by its large mass, is expected to play an 
important role in many new physics models, we only focus, in the following,
on the description of an excited spin-$\tha$ top sector. 

The dynamics of such a top quark excitation,
represented by a four-component Rarita-Schwinger field $\Psi_{\mu}$ of mass $M$, is
described by supplementing to the Standard Model Lagrangian ${\cal L}_{\rm SM}$
the gauge-covariant
version of the Lagrangian of Eq.~\eqref{eq:lageps},
\be
  {\cal L} = {\cal L}_{\rm SM} + \epsilon^{\mu\nu\rho\sigma} \bar{\Psi}_\mu
   \gamma_5 \gamma_\sigma D_\nu\Psi_\rho
     + 2 i M \Psibar_\mu\gamma^{\mu\nu}\Psi_\nu \ ,
\label{eq:ltop}\ee
the spacetime derivative having been replaced by the gauge covariant derivative
\be
  D_\mu\Psi_\nu = \partial_\mu\Psi_\nu - i g_s T_a \Psi_\nu g_\mu^a \ .
\ee
In this expression, we introduce the QCD coupling constant ($g_s$) and the 
fundamental representation matrices of $SU(3)$ ($T_a$).
We also denote the gluon field by
$g_\mu^a$, as in Section~\ref{sec:gravitino}.
Mixing of the spin-$\tha$ top excitation in general occurs through dimension-five
operators suppressed by a new physics energy scale $\Lambda$. Although there are 
several ways to describe such a mixing, we assume it, for the sake of the example,
agnostic of the left-handedness or
right-handedness of the spin-$\oh$ top quark $t$ and embed the relevant interactions within the
Lagrangian
\be\label{eq:ltop2}
  {\cal L}_5 = i\frac{g_s}{\Lambda} \bar\Psi_\rho \Big[ \eta^{\rho\mu} + z \gamma^\rho\gamma^\mu\Big]
    \gamma^\nu T_a t\ g_{\mu\nu}^a  + {\rm h.c.} \  ,
\ee
We leave, in the following,
the off-shell parameter $z$ free\footnote{This parameter gets its name as
it only affects processes with an off-shell spin-$\tha$ field.}.

\begin{figure}
\center
 \includegraphics[width=0.9\columnwidth]{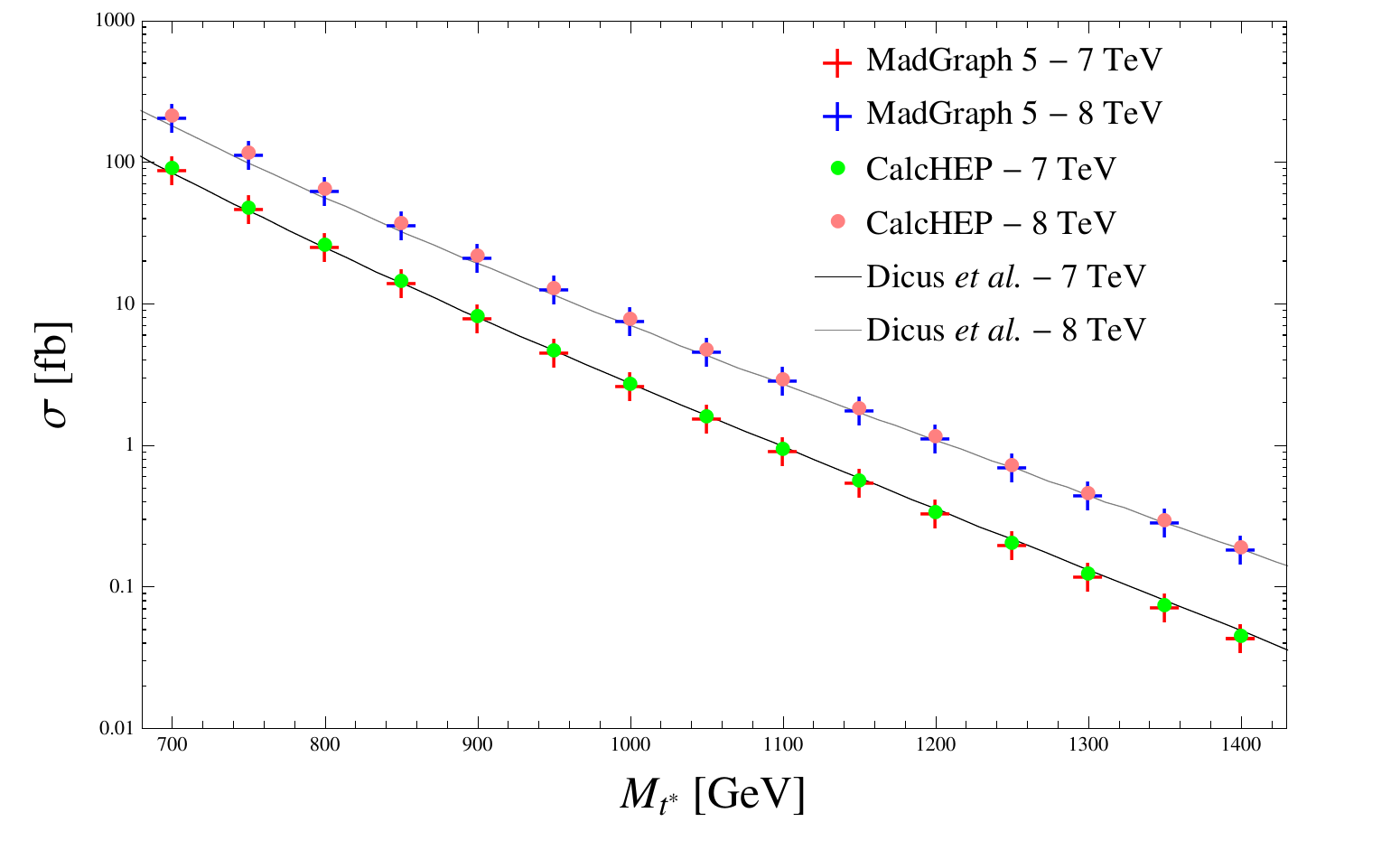}
\caption{Total cross section for the production of an excited top quark pair at the LHC,
  for a center-of-mass energy of 7~TeV (red crosses and green circles) and 8~TeV (blue crosses and red circles),
  as a function of the excited top mass $M_{t^*}$. We compare predictions
  obtained by means of {\sc MadGraph}~5 (crosses) and {\sc CalcHep} (circles)
  to the analytical formulas presented in
  Ref.~\cite{Dicus:2012uh} (lines).
}
\label{fig:tot}
\end{figure}

\subsubsection{Top-quark excitation pair-production}

In order to validate our implementation in the simulation chain mentioned in
Section~\ref{sec:implement}, we first focus on the pair production of spin-$\tha$ top excitations
at the LHC, vetoing diagrams involving a top quark. We compare in
Figure~\ref{fig:tot} predictions derived from the
analytical formulas presented in Ref.~\cite{Dicus:2012uh} to
results obtained by means of the {\sc MadGraph}~5
and {\sc CalcHep} event generators. The relevant model files have been produced by
implementing the Lagrangians of Eqs.~\eqref{eq:ltop}
and \eqref{eq:ltop2} into {\sc FeynRules} and exporting the associated Feynman rules
to a UFO library to be used with {\sc MadGraph}~5 and to a {\sc CalcHep}
model.
We consider the LHC collider
running at a center-of-mass energy of 7~TeV and 8~TeV and convolve
the associated squared matrix element with the leading order fit of the
CTEQ6 parton densities \cite{Pumplin:2002vw} after setting both unphysical
factorization and renormalization scales to the mass of the top excitation $M_{t^*}$.
All calculations fully agree, and the usual behavior of a cross section smoothly falling
with an increasing top-$\tha$ mass is observed. 
From these results, it is found that
spin-$\tha$ excitations of the top
quark with masses $M_{t^*} \lesssim 1$~TeV could have been copiously pair-produced at
both past LHC runs.
Additionally, we have compared the symbolic calculations performed
by {\sc CalcHEP} to the explicit analytic formulas given in Ref.~\cite{Dicus:2012uh} and found perfect agreement.

\begin{figure}
\center
 \includegraphics[width=0.9\columnwidth]{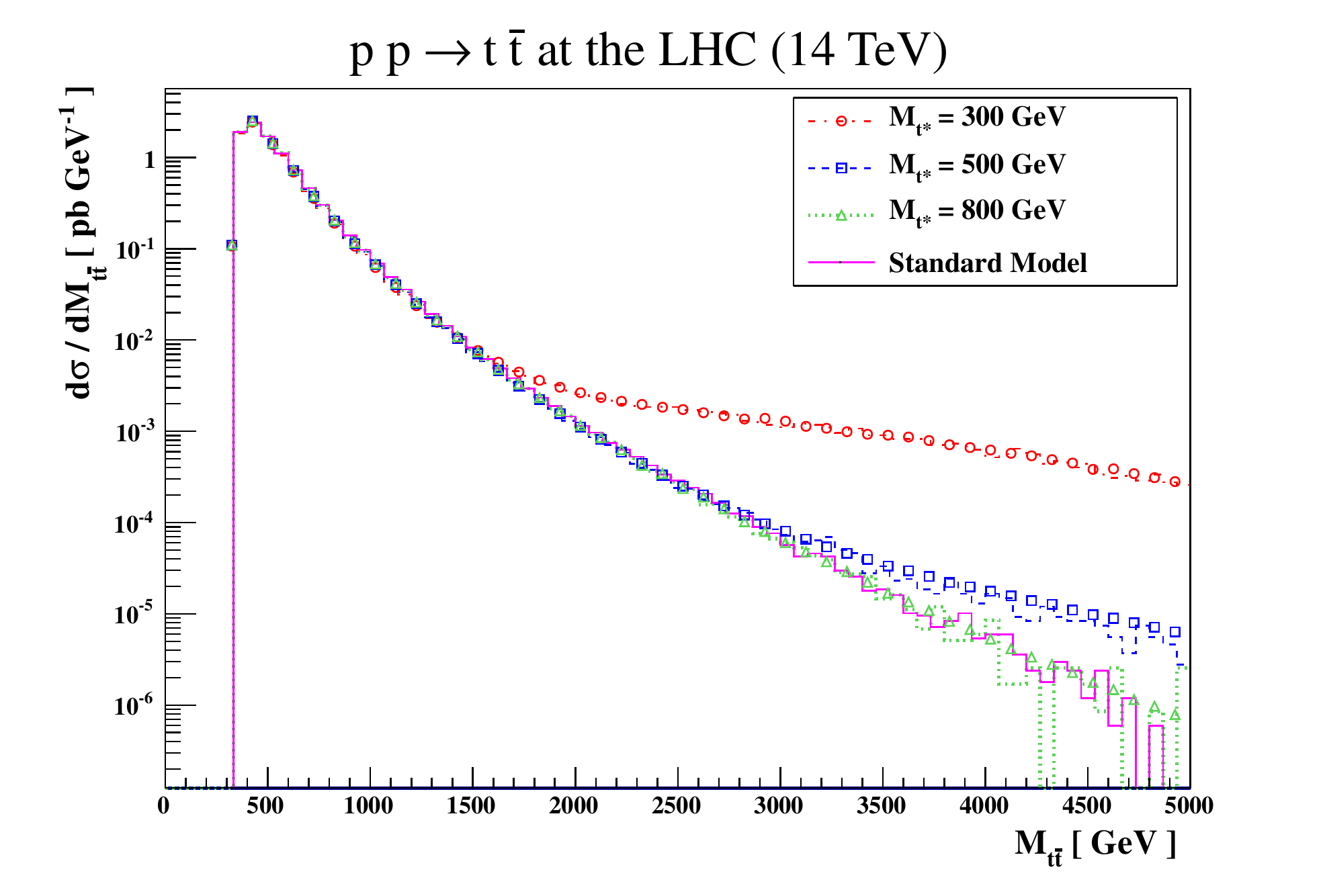}
\caption{Invariant-mass spectrum of a top-antitop pair as would be produced at the LHC, running at
  a center-of-mass energy of 14 TeV, for an excited top mass of $M_{t^*} = 300$~GeV (red dashed-dotted line),
  500~GeV (blue dashed line) and 800~GeV (green dotted line) and in the
  context of the Standard Model (plain purple line).  The dashed lines were obtained from {\sc MadGraph}~5 while the circles, boxes and triangles were obtained from {\sc CalcHEP}.
}
\label{fig:mtt-a}
\end{figure}

\subsubsection{Top-quark pair-production}

Next, mixing effects of spin-$\oh$ and spin-$\tha$ top states are investigated in the
framework of the production of a (spin-$\oh$) top-antitop quark pair at the LHC, running at a center-of-mass
energy of 14 TeV. In the presence of new physics as described by the Lagrangians of Eqs.~\eqref{eq:ltop}
and \eqref{eq:ltop2}, two additional $t$-channel and $u$-channel diagrams lead to the production
of a top-antitop final state via the exchange of a spin-$\tha$ top partner from a gluon-gluon initial state.
Fixing first the off-shell parameter to $z=0$ and the new physics scale to
$\Lambda = 7 M_{t^*}$, we study the variation of the differential
cross section $\d\sigma/\d M_{t\bar t}$ with the mass of the spin-$\tha$ excitation $M_{t^*}$ in
Figure~\ref{fig:mtt-a}. Generating events with
the {\sc MadGraph}~5 program after fixing both unphysical scales to the top mass $M_t = 173$~GeV,
differential distributions are then extracted
with the {\sc MadAnalysis}~5 package \cite{Conte:2012fm} and compared to those
generated using the {\sc CalcHEP} package and its internal histogramming routine.
We recover earlier results \cite{Stirling:2011ya}
and show excesses with respect to the pure Standard Model
case at large top-antitop invariant masses. For the three
scenarios with a respective excited top mass of 300~GeV, 500~GeV and 800~GeV, respectively,
new physics effects appear once the effective scale
$\Lambda$ threshold is crossed, \ie, where the theory becomes unreliable and unitarity is violated.
A proper treatment would require, \eg, the introduction of form factors as in Ref.~\cite{Moussallam:1989nm}.
This however goes beyond the scope of this work devoted to an illustration of
the implementation of spin-$\tha$ fields in automated high-energy physics tools.

\begin{figure}
\center
 \includegraphics[width=0.9\columnwidth]{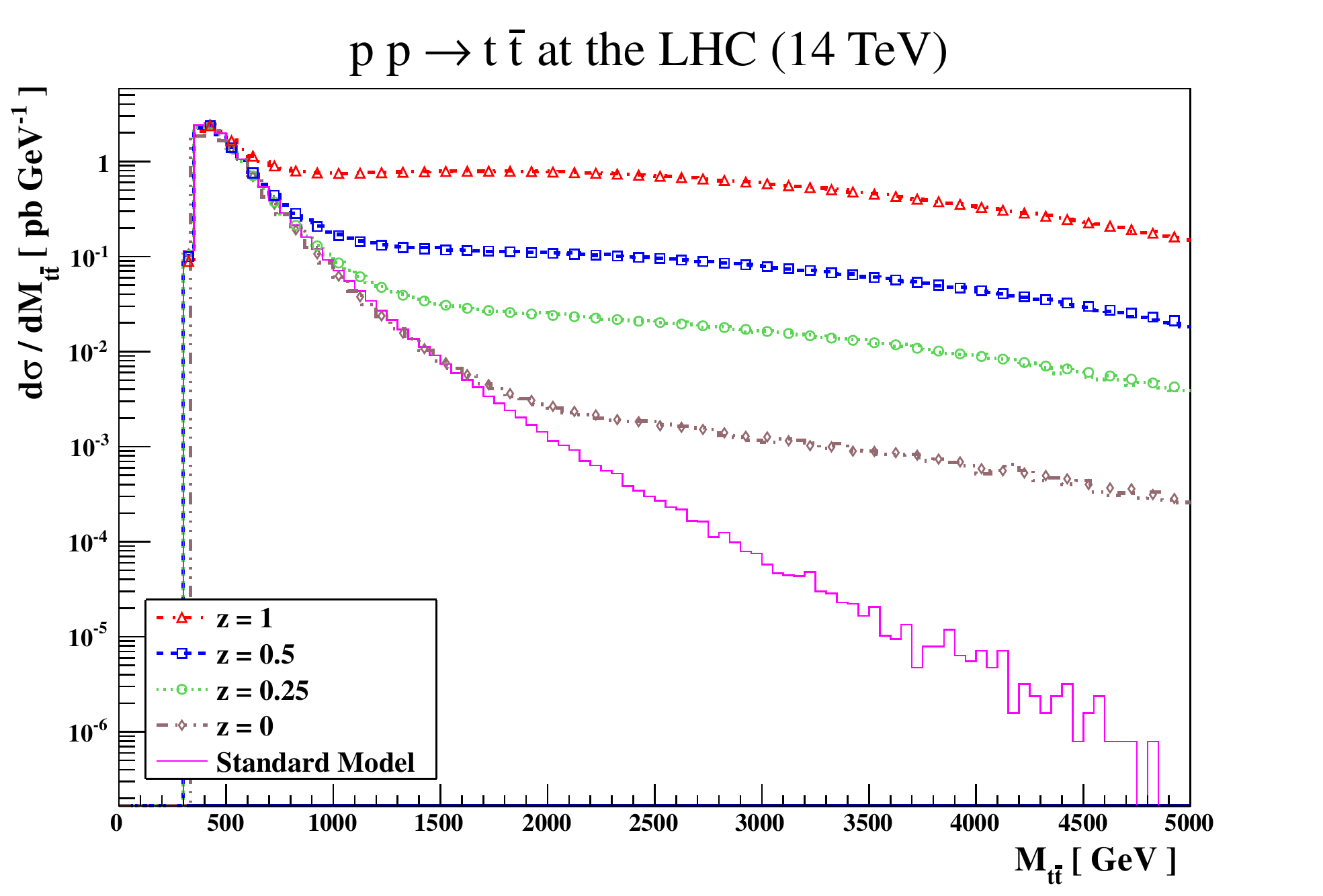}
\caption{Invariant-mass spectrum of a top-antitop pair as would be produced at the LHC, running at
  a center-of-mass energy of 14~TeV, for an excited top mass of $M_{t^*} = 300$~GeV and an off-shell parameter
  fixed to $z=0$ (gray large dashed-dotted line), $z=0.25$ (green dotted line), $z=0.5$ (blue dashed line) and
  \mbox{$z=1$} (red small dashed-dotted line).
  The results are confronted to the Standard Model predictions (plain purple line).}
\label{fig:mtt-b}
\end{figure}

In Figure~\ref{fig:mtt-b}, we fix the top-excitation mass
to $300$~GeV and study the importance of the $z$ parameter on the differential
distribution $\d\sigma/\d M_{t\bar t}$ at the LHC, still assumed to be running at a center-of-mass
energy of 14~TeV. Similarly to the findings of earlier works \cite{Stirling:2011ya},
the off-shell parameter is found to largely control the shape of the $M_{t\bar t}$ distribution. For large positive
$z$-values, new physics effects appear to be important even far below the effective scale $\Lambda$,
in contrast to more popular choices where $z$ is taken vanishing or negative
\cite{Kuhn:1984rj,Moussallam:1989nm}.

%% file: wigner.tex
In this section, we study a particular case of spin-$\tha$
colored resonances that can be thought of as excited quarks. We
exploit our implementation to test
spin correlation effects possibly due to such resonances. In the rest
of this section, we use the notation $u^{*\mu}$ for Rarita-Schwinger fields describing excited spin-$\tha$ quarks
with the same internal quantum numbers as the up quark, and $d^{*\mu}
$ for Rarita-Schwinger fields with the same quantum numbers as the
down quark. If these resonances existed, they could be produced in a
proton-proton collider when one quark and one gluon scatter. The
simplest gauge-invariant operator that can describe such a process is
\be\label{eq:L production}
  {\cal L}_{\text{prod}} = \frac{g_s} {\Lambda} g^a_{\mu \nu} \overline{q} \gamma^{\mu} 
    (k_- P_L + k_+ P_R) T_a q^{* \nu} + {\rm h.c.} \ .
\ee
In this expression, the excited and usual quark fields are denoted by
\mbox{$q^*=(u^*,d^*)$} and \mbox{$q=(u,d)$},
$P_L$ and $P_R$ are chirality projectors acting on spin space,
$\Lambda$ is a cut-off scale and the interaction strengths are described by
the $k_{\pm}$ parameters.
We recall that the objects related to strong interactions have been
introduced in the two previous subsections.
We assume now that the resonances decay into a quark and either a photon or a $Z$ boson. After electroweak symmetry breaking, the simplest Lagrangian describing these decays reads
\be\label{eq:L decay}\bsp
    {\cal L}_{\text{decay}} =&\ \frac{e}{\Lambda} F_{\mu \nu} \overline{q}
       (k_{\gamma-} P_L + k_{\gamma+} P_R) \gamma^\mu q^{* \nu} + {\rm h.c.} \\
   &\ +  \dfrac{g}{\cos\theta_W} Z_\mu \overline{q} (k_{Z-} P_L + k_{Z+} P_R) q^{*\mu} + {\rm h.c.} \ ,
\esp\ee
where we denote the photon field strength tensor by $F_{\mu\nu}$ and model the couplings by
the parameters $k_{\gamma\pm}$ and $k_{Z\pm}$.
In order to study the angular distribution in the decay products caused by these resonances, we
calculate analytically the corresponding differential cross
sections.  For the process $gq\to q^*\to \gamma q$, we find
\be\label{eq:dcs photon}\bsp
\frac{\d\sigma}{\d\Omega} \propto&\ 
   \left(|k_+|^2|k_{\gamma+}|^2+ |k_-|^2|k_{\gamma-}|^2\right)
   \dth{\frac{3}{2}}{\frac{3}{2}}\\
&\ + \left(|k_-|^2|k_{\gamma+}|^2+ |k_+|^2|k_{\gamma-}|^2\right)
\dth{\frac{3}{2}}{-\frac{3}{2}} \ ,
\esp\ee
while for the process $gq\to q^*\to Zq$, we have
\be\label{eq:dcs Z}\bsp
\frac{\d\sigma}{\d\Omega} \propto&\
\left(|k_+k_{Z+}|^2+
  |k_-k_{Z-}|^2\right)\dth{\frac{3}{2}}{\frac{1}{2}}  \\
&\ +
  \left(|k_+k_{Z-}|^2+|k_-k_{Z+}|^2\right)\dth{\frac{3}{2}}{-\frac{1}{2}}\\
&\ +\mathcal{O}\left(\frac{m_Z^2}{m^2}\right)\ .
\esp\ee
For these expressions, we have assumed that the energy in the center-of-mass
frame equals the mass of the resonance $m$.  That is, the spin-$\tha$ particle is on-shell.
We see that expressing the
scattering amplitudes in terms of the Wigner $d$-functions makes the
angular dependence of our 
results more transparent.
We have reviewed the Wigner $d$-functions in Appendix~\ref{sec:Wigner d-functions} and
give the relevant spin-$\tha$ Wigner $d$-functions in Table~\ref{tabledj}.

\renewcommand{\arraystretch}{1.5}%
\begin{table}
	\centering
		\begin{tabular}{c|c|c} \hline\hline
$m$ & $m'$ & $ d^\tha_{mm'}(\theta)$ \\\hline
 $\frac{1}{2}$ & $\frac{1}{2}$  & $\frac{1}{2}\left(3\cos\theta-1\right)\cos\frac{\theta}{2}$ \\\hline
 $\frac{3}{2}$ & $\frac{1}{2}$  & $-\frac{\sqrt{3}}{2}\left(\cos\theta+1\right)\sin\frac{\theta}{2}$ \\\hline
 $\frac{3}{2}$ & $\frac{3}{2}$ & $\frac{1}{2}\left(\cos\theta+1\right)\cos\frac{\theta}{2}$\\\hline\hline
	\end{tabular}
	\caption{\label{tabledj}Wigner $d$-functions used in this paper. The expressions that are not listed are related to the ones shown in the table by means of Eq.~(\ref{reldj}).    }
\end{table}
\renewcommand{\arraystretch}{1}%

\begin{figure}
\begin{center}
\includegraphics[scale=0.70]{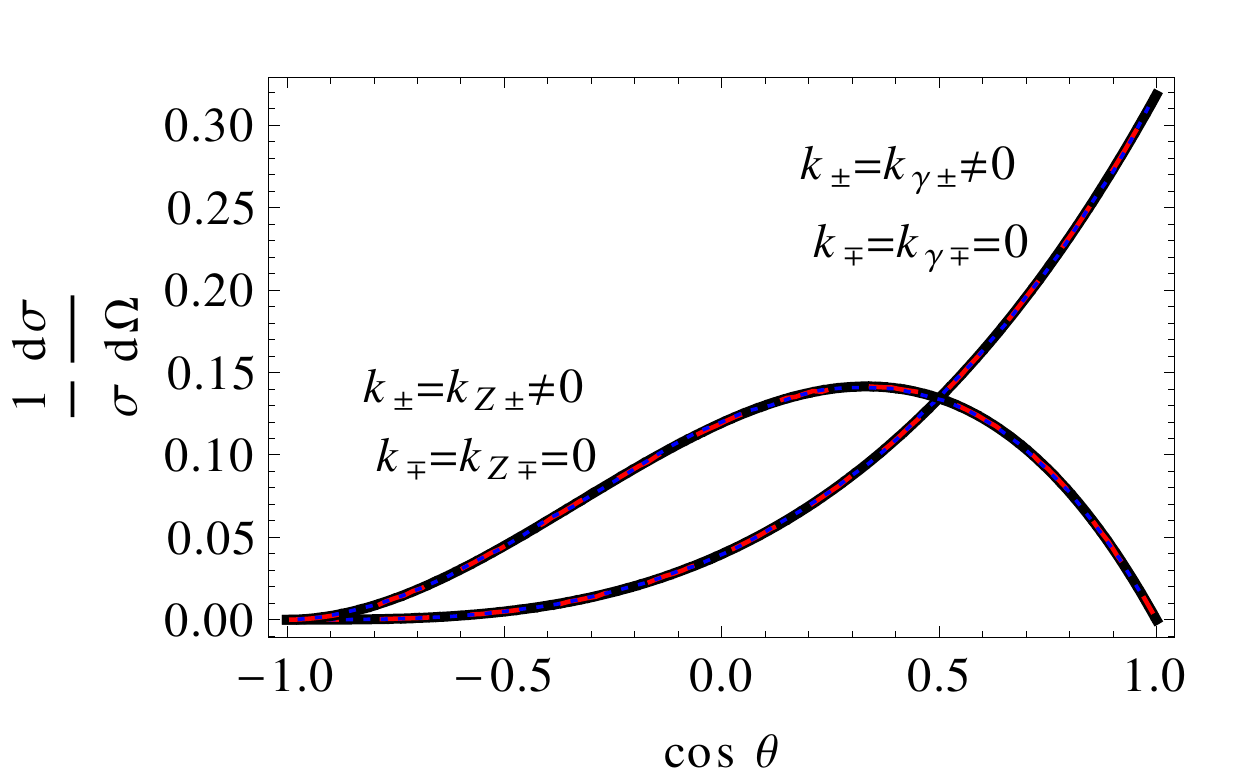}\\
\includegraphics[scale=0.70]{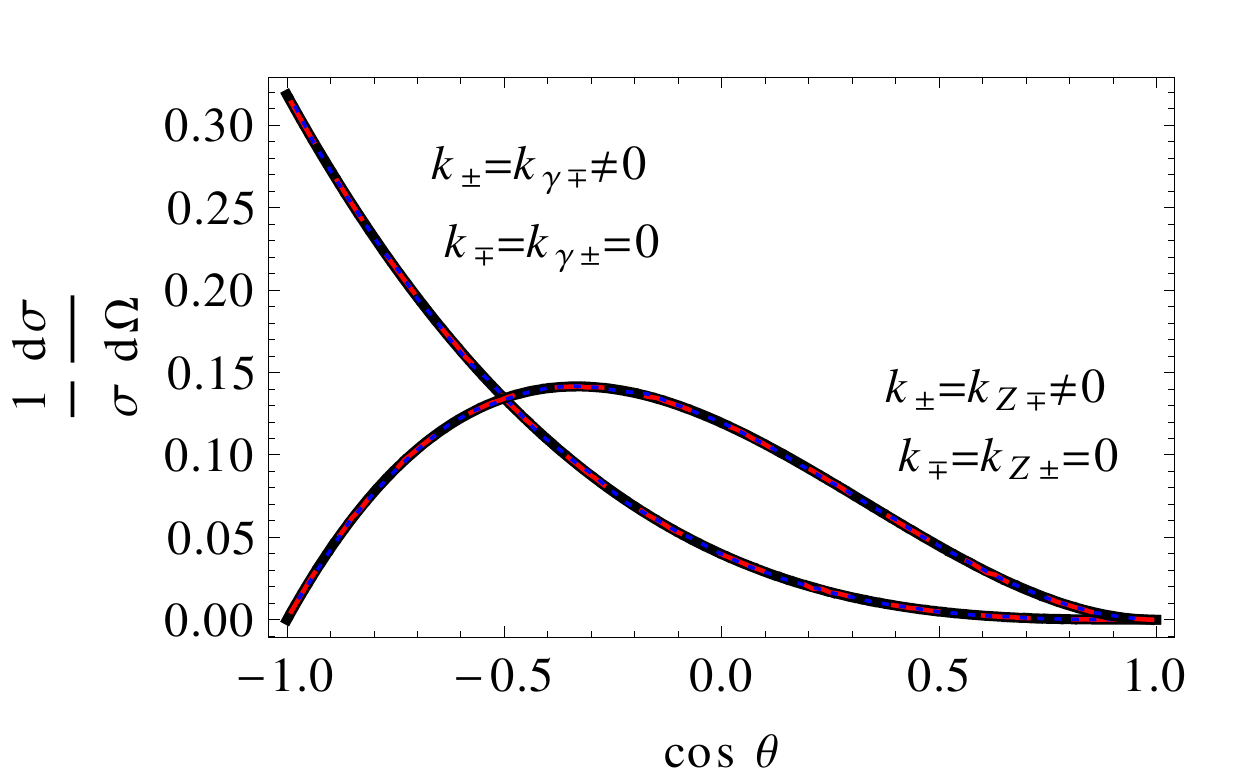}\\
\includegraphics[scale=0.70]{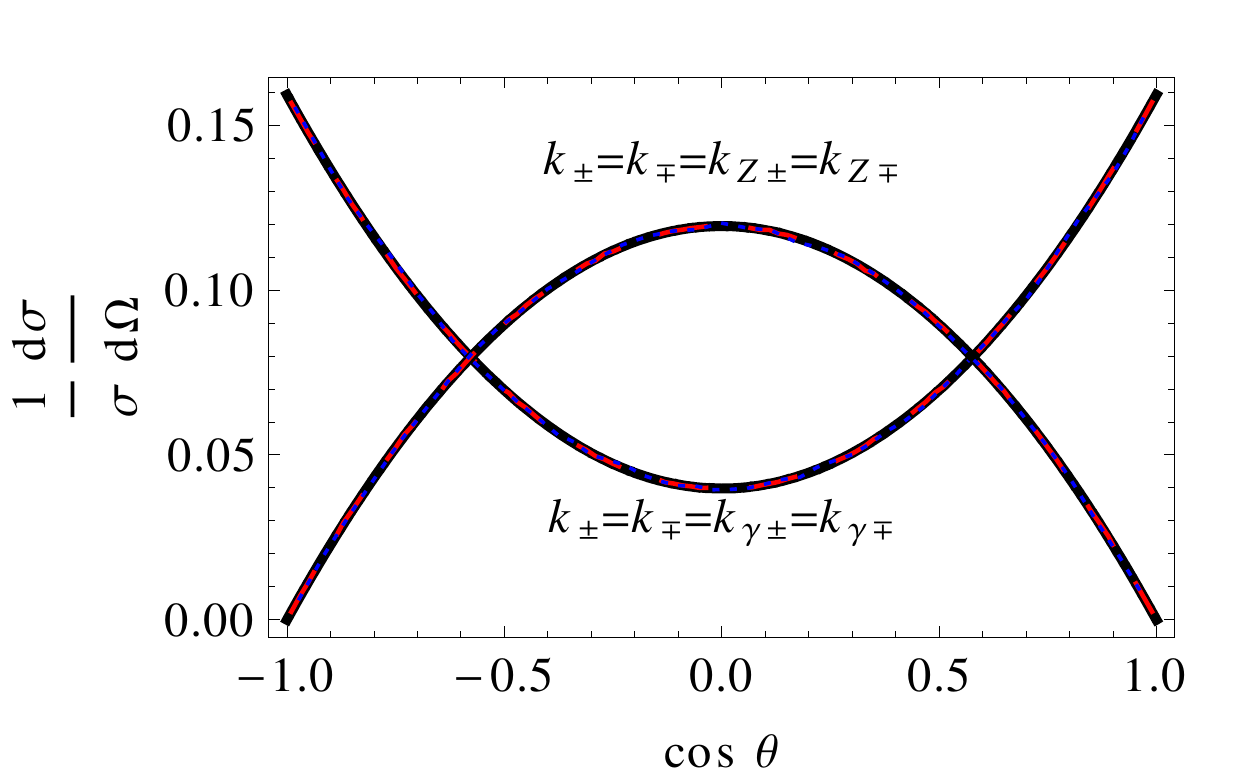}\\
\end{center}
\caption{\label{fig: dist} Angular distributions for three different
choices of couplings. We show analytical results obtained from Eq.~\eqref{eq:dcs photon} and
Eq.~\eqref{eq:dcs Z} (solid black) and numerical predictions of
{\sc CalcHEP} (dashed red) and {\sc MadGraph}~5 (blue dotted).}
\end{figure}

We have implemented this new spin-$\tha$ particle along with the
Lagrangian of Eq.~\eqref{eq:L production} and Eq.~\eqref{eq:L decay} into {\sc FeynRules} and exported the
model to {\sc CalcHEP} and {\sc MadGraph}~5 using the interfaces described in the
previous section.  We have then generated events for the
processes $gq\to q^*\to \gamma q$ and $g q\to q^*\to Z q$ for three
different choices of couplings and show
the resulting distributions in the three panels of
Figure~\ref{fig: dist}. The solid black curves correspond to the
analytical computations of Eq.~\eqref{eq:dcs photon} and Eq.~\eqref{eq:dcs
  Z}, while the red dashed and blue dotted predictions have been numerically obtained by using {\sc CalcHEP}
and {\sc MadGraph}~5, respectively.
  In all cases, the results match perfectly,
illustrating the on-shell propagation of a spin-$\tha$ particle.  For
this study, we used a new physics mass of 1 TeV and width of 1 GeV for the sake of the example,
such a small relative width making the off-shell effects
negligible.

%% file: summary.tex
In this paper, we have reported the implementation of support for spin-$\tha$ fields in the high-energy
physics programs {\sc A\-LO\-HA}, {\sc CalcHEP}, {\sc FeynRules} and {\sc MadGraph}~5, including the
definition of the conventions for those fields in the UFO format.

We have used the chain of packages above-mentioned for several physics applications. First, we have implemented
the gravitino of supergravity together with its interactions in {\sc FeynRules}
and analyzed both gravitino pair-production
and gravitino contributions to gluino pair-production at the LHC using {\sc MadGraph}~5. Our numerical results
have been compared to known analytic expressions that can be found in the literature and 
perfect agreement has been
found. Next, we have implemented a spin-$\tha$ top-quark excitation into {\sc FeynRules}
and studied its pair production using both {\sc MadGraph}~5 and {\sc CalcHEP}.
On the one hand, we have compared the resulting
symbolic calculation of {\sc CalcHEP} with previous analytic calculations. On the other hand,
the numerical results obtained from both {\sc MadGraph}~5 and {\sc Calc\-HEP} have been
confronted to the literature and we have studied the effects implied by the existence of
a top spin-$\tha$ excitation on
top-quark pair production. In all cases, we have found again a perfect agreement.
Finally, we have implemented
an effective operator for spin-$\tha$ colored resonances in {\sc FeynRules},
analyzed the angular distributions in their $s$-channel production mechanism
in both {\sc MadGraph}~5 and {\sc CalcHEP} and compared with analytic calculations involving the Wigner
$d$-functions. Once again, the results have been found to match perfectly.

%% file: conventions.tex
In this Appendix, we collect the conventions adopted in this manuscript.
The metric is given by  
\be 
  \eta_{\mu \nu} = \mathrm{diag}(1,-1,-1,-1) \ , 
\ee 
and the fully antisymmetric tensor of rank four is defined by $\epsilon_{0123}=1$.
The four-vectors built upon the Pauli matrices are given by their usual form,
\be
  \sigma^\mu = \big( 1, \sigma^i \big) \quad\text{and}\quad
  \bar \sigma^\mu = \big(1, -\sigma^i \big) \ , 
\ee
where the Pauli matrices $\sigma^i$ with $i=1,2,3$ read
\be
\sigma^1 = \bpm 0 &  1\\ 1 &  0 \epm\ , \quad 
\sigma^2 = \bpm 0 & -i\\ i &  0 \epm\ , \quad 
\sigma^3 = \bpm 1 &  0\\ 0 & -1 \epm\ .
\ee
This allows to write the generators of the Lorentz algebra in the (two-component)
left-handed and right-handed spinorial representations as
\be
  \sigma^{\mu \nu} = \frac{i}{4} \Big( 
     \sigma^\mu\sibar^\nu - \sigma^\nu\sibar^\mu
  \Big)
   \quad \text{and} \quad 
  \sibar^{\mu \nu} = \frac{i}{4}\Big( 
     \sibar^\mu \sigma^\nu - \sibar^\nu \sigma^\mu
  \Big) \ ,
\ee
respectively.

Moving to four-component spinors, Dirac matrices are defined in the Weyl representation by
\be
\gamma^\mu=\bpm 0& \sigma^\mu\\
\bar \sigma^\mu&0 \epm\ ,
\ee
and span the Clifford algebra
\be
 \Big\{ \gamma^\mu, \gamma^\nu\Big\} = 2 \eta^{\mu\nu} \ .
\label{eq:cliff}\ee
Additionally, the fifth Dirac matrix $\gamma_5$ is given by
\be
\gamma_5 = i\gamma^0 \gamma^1\gamma^2 \gamma^3 = \bpm -1&0\\0&1 \epm \ .
\ee
The $\gamma$-matrices allow to build the generators of the Lorentz algebra in the
four-component spinorial representation,
\be
 \gamma^{\mu\nu} = \frac{i}{4} \Big(\gamma^\mu\gamma^\nu - \gamma^\nu\gamma^\mu\Big)\ ,
\ee
as well as the quantity $\gamma^{\mu\nu\rho}$ which may appear in some specific forms
of the Rarita-Schwinger Lagrangian used in the literature,
\be\bsp
 \gamma^{\mu\nu\rho} =&\ \frac16 \Big[ \gamma^\mu\gamma^\nu\gamma^\rho + 
      \gamma^\nu\gamma^\rho\gamma^\mu + \gamma^\rho\gamma^\mu\gamma^\nu
 \\ &\ \quad - \gamma^\nu\gamma^\mu\gamma^\rho - \gamma^\mu\gamma^\rho\gamma^\nu
   - \gamma^\rho\gamma^\nu\gamma^\mu \Big] \ .
\esp\ee
The $\gamma^{\mu\nu}$ and $\gamma^{\mu\nu\rho}$ objects obey the important relations,
\be
   \epsilon^{\mu\nu\rho\sigma} \gamma_5 \gamma_\sigma =   -i \gamma^{\mu\nu\rho}  \quad\text{and}\quad
   \epsilon^{\mu\nu\rho\sigma} \gamma_{\mu\nu} \gamma_5 =  2 i \gamma^{\rho\sigma}  \ ,
\label{eq:Diracprop1}\ee
helpful to prove the equivalence between all the spin-$\tha$ Lagrangian forms employed up to now.

%% file: polarization_app.tex
In this appendix, we follow the approach of Weinberg to build up the properties of a spin-$\tha$
particle~\cite{Weinberg:1995mt}.  The spin of a particle determines its polarization vectors,
the latter determining its propagation on-shell.  Off-shell, the propagator can have further
dependence which vanishes on-shell. Moreover, the on-shell propagator is related
to the on-shell quadratic part of the Lagrangian whereas  the off-shell pieces of the propagator
match the model-dependent off-shell quadratic terms of the Lagrangian.  We begin with a discussion on spin.

Both the vacuum and the dynamics of quantum field theory are symmetric under the Lorentz group,
whose algebra $\mathfrak{so}(3,1)$ is isomorphic to
$\mathfrak{sl}(2,\mathbb R) \oplus\overline{\mathfrak{sl}(2,\mathbb R)}$, the representations of the two
$\mathfrak{sl}(2,\mathbb R)$ denoting the left-handed and right-handed chiral algebras.
However, the four-momentum of the particles are further invariant under a subalgebra, called the little algebra.  For a massive particle, this is easiest to see in the rest frame where its momentum is $p^\mu=(M,0,0,0)$.  Any Lorentz transformation that is a pure rotation in space leaves invariant a massive particle at rest.  As a result, the little algebra for a massive particle is $\mathfrak{so}(3)$, the algebra of spatial rotations, which is isomorphic to
$\mathfrak{su}(2)$, and each massive particle is classified according to its transformation properties under this little algebra or, in other words, according to its ``spin".
This little algebra is the diagonal subalgebra of the left and right chiral Lorentz algebras,
$\mathfrak{sl}(2,\mathbb R)$ and $\overline{\mathfrak{sl}(2,\mathbb R)}$.
As a result, if a field transforms under the $(a,b)$ representation of the Lorentz algebra,
it has a spin between $a+b$ and $|a-b|$.  

As is well known, a scalar field transforms as $(0,0)$ under the Lorentz algebra and is, therefore, a spin-$0$ field.  Ordinary matter fermions transform as either $(\oh,0)$ or $(0,\oh)$ and are, therefore, spin-$\oh$ particle.
Vector fields transform as $(\oh,\oh)$ and contain both a spin-$1$ and a spin-$0$ piece.  On-shell, the spin-$0$ field is removed
by $\partial_\mu V^\mu=0$.  Rarita-Schwinger fields $\Psi^\mu$ are formed as a direct product of $(\oh,0)$ or $(0,\oh)$ and $(\oh,\oh)$.
Under the Lorentz symmetry, this contains
\begin{itemize}
 \item $(0,\oh)$ or $(\oh,0)$ parts which are spin-$\oh$ fields and are removed by the on-shell equation $\gamma_\mu \Psi^\mu=0$;
  \item $(1,\oh)$ or $(\oh,1)$ parts which contain both the spin-$\tha$ field that we are after and another spin-$\oh$ field which is removed by the on-shell equation $\partial_\mu\Psi^\mu=0$.
\end{itemize}
Higher spin states are formed in a similar way.

In the rest of this appendix, we review the spin algebra $\mathfrak{su}(2)$ (Section~\ref{sec: spin review}),
the polarization vectors (Section~\ref{sec: spin polarization vectors}), the propagators
(Section~\ref{sec:propagator}), the quadratic Lagrangian terms (Section~\ref{sec: Lagrangian}), and finally
the Wigner $d$-functions (Section~\ref{sec:Wigner d-functions}).  Further, for the rest of this Appendix, we
consider particles which are not self-charge-conjugate. The special case when particles are self-conjugate
can be recovered from our
results by replacing antiparticle states by charge conjugates of the particle states.

\subsection{\label{sec: spin review}Spin: $\mathbf{\mathfrak{so(3)}\sim \mathfrak{su(2)}}$}

We begin by reviewing what it means to be a representation of the $\mathfrak{su}(2)$ spin algebra.  By definition, in this algebra there are three generators that satisfy the commutator rule
\begin{equation}\label{eq:su2 structure constant}
\left[ J_i , J_j \right] = i \epsilon_{ijk} J_k \ .
\end{equation}
Two of these generators can be combined to form the raising/lowering operators,
\begin{equation}
J_\pm = J_1 \pm i J_2 \ ,
\end{equation}
which fulfill the commutator rules
\be
\left[J_+ , J_-\right] = 2 J_3
\quad\text{and}\quad
\left[J_3 , J_\pm\right] = \pm J_\pm \ .
\ee
These can easily be checked by means of Eq.~(\ref{eq:su2 structure constant}).  With
these definitions, it is easy to show that the effects of the rai\-sing/lo\-we\-ring operators on
a state of spin $j$ represented by $|j,\sigma\rangle$ is to
increase/decrease the eigenvalue of $J_3$ denoted by $\sigma$,
\begin{equation}
J_3 J_\pm |j,\sigma\rangle = (\sigma\pm1) J_\pm |j,\sigma\rangle \ .
\end{equation}

A couple of general remarks are in order before we consider specific spin. 
Given the above definitions, we can easily see that the set of generators given by flipping the signs of all the generators and by switching $J_+$ with $J_-$ also forms a representation of the same algebra.  We will show this by defining
\be
\bar{J}_3 = - J_3
\quad\text{and}\quad
\bar{J}_\pm = - J_\mp \ .
\ee
With these, we easily obtain
\be
\left[ \bar{J}_+ , \bar{J}_- \right] = 2 \bar{J}_3
\quad\text{and}\quad
\left[ \bar{J}_3 , \bar{J}_\pm \right] = \pm \bar{J}_\pm \ .
\ee
This representation is called the complex representation and is very important in what follows. It finds
its name due to the property
\begin{equation}
\left[ -J^*_i , -J^*_j \right] = i \epsilon_{ijk} \left(-J^*_k\right)\ .
\end{equation}
Since we are taking our states as eigenvectors of $J_3$, this representation
is diagonal and real (the eigenvalues of a Hermitian operator are real), so that
$\bar{J}_3=-J_3^*=-J_3$.  Further, in this basis, $J_1$ is real and symmetric while $J_2$ is imaginary and antisymmetric, so $\bar{J}_1=-J^*_1=-J_1$ and $\bar{J}_2=-J^*_2=J_2$.  As a result, $\bar{J}_\pm=\left(-J^*_1\right)\pm i\left(-J^*_2\right)=-\left(J_1\mp i J_2\right)=-J_\mp$.

Moreover, these raising/lowering operators can always be shifted by a phase and still satisfy the algebra.  For example, if we define
\begin{equation}
J_\pm^\prime = e^{\pm i\phi}J_\pm \ ,
\end{equation}
it is easy to show that
\be
\left[ J_+^\prime , J_-^\prime \right] = 2 J_3
\quad\text{and}\quad
\left[ J_3 , J_\pm^\prime \right] = \pm J_\pm^\prime \ .
\ee
So that the set $J_3$ with $J_\pm^\prime$ also forms a representation of $\mathfrak{su}(2)$.
This will be important for understanding the {\sc Helas} conventions for the spin-$\oh$ polarization vectors.

For the rest of this section, we consider the particle or antiparticle to have a mass $M$ and a four-momentum
\begin{equation}
p^\mu = \left(E,|\vec{p}|\sin\theta\cos\phi,|\vec{p}|\sin\theta\sin\phi,|\vec{p}|\cos\theta\right)
\end{equation}

\subsection{\label{sec: spin polarization vectors}Polarization vectors}
The polarization vectors for a particle $u$ and its antiparticle $v$ are determined by the spin of the particle.   They transform under the spin representation of the particle and antiparticle, respectively, which is to say, the particle's spin transforms according to $\vec{J}$ while the antiparticle's spin transforms according to $-\vec{J}^*$.  On the other hand, they both transform according to the same representation of the Lorentz algebra whose generators are $\vec{\mathcal{J}}$.  For the rest of this discussion, we will take $J_3$ to measure the component of spin along the direction of motion.  In other words, we will take $J_3$ to be the helicity operator.  $J_1$ and $J_2$ will be defined appropriately, with the direction $1$ and $2$ orthogonal to direction $3$ and each other and satisfying Eq.~\eqref{eq:su2 structure constant}.  

As we mentioned, the spin algebra is a subalgebra of the Lorentz algebra.
So, each spin transformation corresponds with a Lorentz transformation.  For particles, we can write this
as\footnote{From now on, we explicitly indicate the spin quantum number $j$ as an upper
index of the $J_3$ operator, \ie, $J_3^j$.}
\be\bsp
J^j_3 u_\ell(j,\sigma) =&\ \mathcal{J}^{(3)}_{\ell,\ell'} u_{\ell'}(j,\sigma) = \sigma u_\ell(j,\sigma) \ , \\
J^j_\pm u_\ell(j,\sigma) =&\ \mathcal{J}^{(\pm)}_{\ell,\ell'} u_{\ell'}(j,\sigma) \\
=&\ \sqrt{(j\mp\sigma)(j\pm\sigma+1)} u_{\ell}(j,\sigma\pm1)\  ,
\esp\ee
where $\ell$ and $\ell'$ are the field indices under the Lorentz transformations and $\ell'$ is summed over.
Furthermore, we have assumed a trivial phase for the raising and lowering operators.  Other phase choices
are discussed later.
Antiparticles transform according to the same representation of the Lorentz algebra and, so, we use
the same $\vec{\mathcal{J}}$ operators.
However, their spin transforms under the conjugate representation.  As a result, the Lorentz transformations that were used for particles have the following effect on antiparticles,
\be\bsp
-\bar{J}^j_3 v_\ell(j,\sigma) =&\ \mathcal{J}^{(3)}_{\ell,\ell'} v_{\ell'}(j,\sigma) = -\sigma v_\ell(j,\sigma)\ ,\\
-\bar{J}^j_\mp v_\ell(j,\sigma) =&\ \mathcal{J}^{(\pm)}_{\ell,\ell'} v_{\ell'}(j,\sigma) \\
=&\ -\sqrt{(j\pm\sigma)(j\mp\sigma+1)} v_{\ell}(j,\sigma\mp1)\ .
\esp\ee

For the rest of this subsection, we review the polarization vectors for spin-$\oh$, spin-$1$, spin-$\tha$, and spin-$2$
fields. We use the {\sc Helas} convention for the spin-$\oh$ and spin-$1$ polarization vectors and show that
they satisfy the spin algebra and derive any relevant phases in the raising and lowering operators.  We then construct the spin-$\tha$ and spin-$2$ polarization vectors from these. 

\subsubsection{\label{sec: spin 1/2 polarization vectors}Spin-$\oh$ fields}
According to the {\sc Helas} conventions, the spin-$\oh$ polarization vectors are given by
\be\bsp
u_{+1/2}(p) =&\
\left(\begin{array}{c}
\sqrt{E-|\vec{p}|}\left(\begin{array}{c}\cos\frac{\theta}{2}\\\sin\frac{\theta}{2}e^{i\phi}\end{array}\right)\\
\sqrt{E+|\vec{p}|}\left(\begin{array}{c}\cos\frac{\theta}{2}\\\sin\frac{\theta}{2}e^{i\phi}\end{array}\right)
\end{array}\right)\ ,\\
u_{-1/2}(p) =&\ 
\left(\begin{array}{c}
\sqrt{E+|\vec{p}|}\left(\begin{array}{c}-\sin\frac{\theta}{2}e^{-i\phi}\\\cos\frac{\theta}{2}\end{array}\right)\\
\sqrt{E-|\vec{p}|}\left(\begin{array}{c}-\sin\frac{\theta}{2}e^{-i\phi}\\\cos\frac{\theta}{2}\end{array}\right)
\end{array}\right)\ ,\\
v_{+1/2}(p) =&\ 
\left(\begin{array}{c}
-\sqrt{E+|\vec{p}|}\left(\begin{array}{c}-\sin\frac{\theta}{2}e^{-i\phi}\\\cos\frac{\theta}{2}\end{array}\right)\\
\sqrt{E-|\vec{p}|}\left(\begin{array}{c}-\sin\frac{\theta}{2}e^{-i\phi}\\\cos\frac{\theta}{2}\end{array}\right)
\end{array}\right)\ ,\\
v_{-1/2}(p) =&\
\left(\begin{array}{c}
\sqrt{E-|\vec{p}|}\left(\begin{array}{c}\cos\frac{\theta}{2}\\\sin\frac{\theta}{2}e^{i\phi}\end{array}\right)\\
-\sqrt{E+|\vec{p}|}\left(\begin{array}{c}\cos\frac{\theta}{2}\\\sin\frac{\theta}{2}e^{i\phi}\end{array}\right)
\end{array}\right)\label{eq:vm} \ .
\esp\ee

We begin with the spin-$\oh$ form of the rotation and boost generators
\be
\mathcal{J}_i = \frac{1}{2}\left(\begin{array}{cc}\sigma_i&0\\0&\sigma_i\end{array}\right) \quad\text{and}\quad
\mathcal{K}_i = -\frac{i}{2}\left(\begin{array}{cc}\sigma_i&0\\0&-\sigma_i\end{array}\right) \ ,
\ee
where $\sigma_i$ denotes the Pauli matrices,
and then combine $\mathcal{J}_1$ and $\mathcal{J}_2$ to form the ladder operators
\begin{equation}
\mathcal{J}_\pm = \mathcal{J}_1\pm i \mathcal{J}_2 =
   \frac{1}{2}\left(\begin{array}{cc}\sigma_1\pm i \sigma_2&0\\0&\sigma_1\pm i \sigma_2\end{array}\right) \ .
\end{equation}
We then rotate and boost from the particle rest frame to the laboratory frame of reference
by employing the operator
\begin{eqnarray}
 &&\Lambda_{\frac{1}{2}}=\\ \nonumber &&\quad
   \left(\begin{array}{cccc}
e^{-\frac{\eta}{2}-i\frac{\phi}{2}}c_{\frac{\theta}{2}}&-e^{\frac{\eta}{2}-i\frac{\phi}{2}}s_{\frac{\theta}{2}}&0&0\\
e^{-\frac{\eta}{2}+i\frac{\phi}{2}}s_{\frac{\theta}{2}}&e^{\frac{\eta}{2}+i\frac{\phi}{2}}c_{\frac{\theta}{2}}&0&0\\
0&0&e^{\frac{\eta}{2}-i\frac{\phi}{2}}c_{\frac{\theta}{2}}&-e^{-\frac{\eta}{2}-i\frac{\phi}{2}}s_{\frac{\theta}{2}}\\
0&0&e^{\frac{\eta}{2}+i\frac{\phi}{2}}s_{\frac{\theta}{2}}&e^{-\frac{\eta}{2}+i\frac{\phi}{2}}c_{\frac{\theta}{2}}
\end{array}\right) \ .
\end{eqnarray}
In this expression, we have introduced the sine and cosine of half the polar angle
$c_{\theta/2} = \cos(\theta/2)$ and $s_{\theta/2} = \sin(\theta/2)$ as well as
the pseudorapidity
\begin{equation}
 \eta = \frac{1}{2}\ln\left(\frac{E+|\vec{p}|}{E-|\vec{p}|}\right) \ .
\end{equation}
At the operator level, we hence get
\be\bsp
\mathcal{J}^{(3)} =&\ \Lambda_{\frac{1}{2}}\mathcal{J}_z\Lambda_{\frac{1}{2}}^{-1} =
\left(\begin{array}{cc}B&0\\0&B\end{array}\right)\ , \\
\mathcal{J}^{(\pm)} =&\ \Lambda_{\frac{1}{2}}\mathcal{J}_\pm\Lambda_{\frac{1}{2}}^{-1} =
\left(\begin{array}{cc}e^{\mp\eta}A_\pm&0\\0&e^{\pm\eta}A_\pm\end{array}\right) \ , 
\esp\ee
with
\be\bsp
A_\pm=&\ \left(\begin{array}{cc}
-\frac{1}{2}\sin\theta&\pm\frac{1}{2}\left(1\pm\cos\theta\right)e^{-i\phi}\\
\mp\frac{1}{2}\left(1\mp\cos\theta\right)e^{i\phi}&\frac{1}{2}\sin\theta
\end{array}\right) \ , \\
B=&\ \left(\begin{array}{cc}\frac{1}{2}\cos\theta&\frac{1}{2}\sin\theta e^{-i\phi}\\\frac{1}{2}\sin\theta e^{i\phi}&-\frac{1}{2}\cos\theta\end{array}\right) \ .
\esp\ee
It can easily be shown that these operators satisfy the commutation properties of $\mathfrak{su}(2)$,
\be
\left[ \mathcal{J}^{(+)} , \mathcal{J}^{(-)} \right] = 2 \mathcal{J}^{(3)}\quad\text{and}\quad
\left[ \mathcal{J}^{(3)} , \mathcal{J}^{(\pm)} \right] = \pm \mathcal{J}^{(\pm)} \ .
\ee
They correspond with the helicity operator, raising operator and lowering operator in spin space for particles.
We can explicitly check that they have the following effect on the particle polarization vectors of Eq.~\eqref{eq:vm},
\be\bsp
J^{1/2}_3 u_{\pm 1/2}(p) =&\ \mathcal{J}^{(3)} u_{\pm1/2}(p) = \pm\frac{1}{2} u_{\pm1/2}(p)\ , \\
J^{1/2}_- u_{1/2}(p) =&\ \mathcal{J}^{(-)} u_{1/2}(p) = e^{i\phi}u_{-1/2}(p) \ , \\
J^{1/2}_- u_{-1/2}(p) =&\ \mathcal{J}^{(-)} u_{-1/2}(p) = 0 \ , \\
J^{1/2}_+ u_{-1/2}(p) =&\ \mathcal{J}^{(+)} u_{-1/2}(p) = e^{-i\phi}u_{1/2}(p)\ , \\
J^{1/2}_+ u_{1/2}(p) =&\ \mathcal{J}^{(+)} u_{1/2}(p) = 0 \ .
\esp\ee
From this, we learn that there is an extra phase associated with the ladder operators in the {\sc Helas}
conventions for spin-$\oh$,
\be
J^{1/2}_\pm u_\sigma(p) = e^{\mp i\phi}\sqrt{\left(\oh\mp\sigma\right)\left(\oh\pm\sigma+1\right)}\ u_{\sigma\pm1}(p) \ .
\ee
Since the generators formed in this way still form a representation of the spin algebra, this phase could have been absorbed into the definition of the polarization vectors.

For antiparticles, we find
\be\bsp
\bar{J}^{1/2}_3 v_{\pm 1/2}(p) =&\ -\mathcal{J}^{(3)} v_{\pm 1/2}(p) = \pm\frac{1}{2}v_{\pm 1/2}(p) \ ,\\
\bar{J}^{1/2}_- v_{1/2}(p) =&\ -\mathcal{J}^{(+)} v_{1/2}(p) = e^{-i\phi}v_{-1/2}(p) \ ,\\
\bar{J}^{1/2}_- v_{-1/2}(p) =&\ -\mathcal{J}^{(+)} v_{-1/2}(p) = 0\ ,\\
\bar{J}^{1/2}_+ v_{-1/2}(p) =&\ -\mathcal{J}^{(-)} v_{-1/2}(p) = e^{i\phi}v_{1/2}(p)\ ,\\
\bar{J}^{1/2}_+ v_{1/2}(p) =&\ -\mathcal{J}^{(-)} v_{1/2}(p) = 0 \ ,
\esp\ee
and as a result, 
\be
\bar{J}^{1/2}_\pm v_\sigma(p) = e^{\pm i\phi}\sqrt{\left(\oh\mp\sigma\right)\left(\oh\pm\sigma+1\right)}\ v_{\sigma\pm1}(p) \ .
\ee

\subsubsection{\label{sec: spin 1 polarization vectors}Spin-$1$ fields}
The polarization vectors associated with a spin-$1$ particle are given, following the
{\sc Helas} conventions, by
\be\bsp
\epsilon^\mu_+(p) =&\
\frac{1}{\sqrt{2}}\left(\begin{array}{c}
0\\
-\cos\theta\cos\phi+i\sin\phi\\
-\cos\theta\sin\phi-i\cos\phi\\
\sin\theta
\end{array}\right)\ , \\
\epsilon^\mu_0(p) =&\
\frac{1}{M}\left(\begin{array}{c}|\vec{p}|\\E\sin\theta\cos\phi\\E\sin\theta\sin\phi\\E\cos\theta\end{array}\right)\ ,\\
\epsilon^\mu_-(p) =&\
\frac{1}{\sqrt{2}}\left(\begin{array}{c}
0\\
\cos\theta\cos\phi+i\sin\phi\\
\cos\theta\sin\phi-i\cos\phi\\
-\sin\theta
\end{array}\right)\label{eq:epm} \ ,
\esp\ee
whereas those associated with a spin-$1$ antiparticle read
\be\bsp
\bar{\epsilon}_+^\mu(p) =&\ \epsilon^{\mu *}_+(p)\ ,\\
\bar{\epsilon}_0^\mu(p) =&\ \epsilon^{\mu *}_0(p)\ ,\\
\bar{\epsilon}_-^\mu(p) =&\ \epsilon^{\mu *}_-(p)\label{eq:epmc} \ .
\esp\ee

In the vectorial representation, the spatial pieces of the rotation generators are given by
\begin{equation}
(\mathcal{J}_i)^j{}_k = i \eps_i{}^j{}_k\ ,
\end{equation}
where $\eps_{ijk} = 1$ and $i$, $j$, $k$ = 1,2,3, while the time components are zero.
On the other hand, the spatial components of the
boost generators vanish while their time components are given by
\be
  (\mathcal{K}_i)^0{}_\mu = i \eta_{i\mu}
  \quad\text{and}\quad
  (\mathcal{K}_i)^\mu{}_0 = -i \delta^\mu{}_i \ .
\ee
As a result, in the rest frame, we have
\be\bsp
\mathcal{J}_3 = \left(\begin{array}{cccc}0&0&0&0\\0&0&-i&0\\0&i&0&0\\0&0&0&0\end{array}\right)\quad\text{and}\quad
\mathcal{J}_\pm =
  \left(\begin{array}{cccc}0&0&0&0\\0&0&0&\mp1\\0&0&0&-i\\0&\pm1&i&0\end{array}\right) \ .
\esp\ee
To get these generators in the laboratory frame of reference, we boost along the $z$-direction and rotate using 
\be
\Lambda_1=
\left(\begin{array}{cccc}1&0&0&0\\0&c_\phi&-s_\phi&0\\0&s_\phi&c_\phi&0\\0&0&0&1\end{array}\right)
\left(\begin{array}{cccc}1&0&0&0\\0&c_\theta&0&s_\theta\\0&0&1&0\\0&-s_\theta&0&c_\theta\end{array}\right)
\frac{1}{M}\left(\begin{array}{cccc}E&0&0&p\\0&1&0&0\\0&0&1&0\\p&0&0&E\end{array}\right) \ ,
\ee
where the sine and cosine of the azimuthal (polar) angle are denoted by $s_\phi$ ($s_\theta$)
and $c_\phi$ ($c_\theta$).
One can now derive the rotation generators
\be\bsp
\mathcal{J}^{(3)} =&\ \Lambda_1 \mathcal{J}_3 \Lambda_1^{-1}
=
\left(\begin{array}{cccc}
0&0&0&0\\
0&0&-ic_\theta & i s_\theta s_\phi\\
0&ic_\theta&0&-is_\theta c_\phi\\
0&-is_\theta s_\phi&is_\theta c_\phi&0
\end{array}\right) \ ,\\
\mathcal{J}^{(\pm)}=&\ \Lambda_1 \mathcal{J}_\pm \Lambda_1^{-1}
\\ =&\ \left(\begin{array}{cccc}
0&
\frac{p}{M}C_{1\pm} &
\frac{p}{M}C_{2\pm} &
\mp\frac{ps_\theta}{M}\\
\frac{p}{M}C_{1\pm} &
0&
\frac{iEs_\theta}{M} &
-\frac{E}{M}C_{3\pm} \\
\frac{p}{M}C_{2\pm} &
-\frac{iEs_\theta}{M} &
0&
-\frac{E}{M}C_{4\pm} \\
\mp\frac{ps_\theta}{M} &
\frac{E}{M}C_{3\pm} &
\frac{E}{M}C_{4\pm} &
0
\end{array}\right) \ ,
\esp\ee
where we have introduced the quantities
\be\bsp
C_{1\pm} =&\ \pm\cos\theta\cos\phi-i\sin\phi\ , \\
C_{2\pm} =&\ \pm\cos\theta\sin\phi+i\cos\phi\ , \\
C_{3\pm} =&\ -i\cos\theta\sin\phi\pm\cos\phi\ ,  \\
C_{4\pm} =&\ i\cos\theta\cos\phi\pm\sin\phi \ .
\esp\ee
The commutators of these generators satisfy the usual relations,
\be
\left[ \mathcal{J}^{(+)} , \mathcal{J}^{(-)} \right] = 2 \mathcal{J}^{(3)}
\quad\text{and}\quad
\left[ \mathcal{J}^{(3)} , \mathcal{J}^{(\pm)} \right] = \pm \mathcal{J}^{(\pm)} \ .
\ee
Acting on the spin-1 polarization vectors, we get
\be\bsp
J_3 \epsilon_\pm^\mu(p) =& \mathcal{J}^{(3)\mu}_{\hspace{0.2in}\nu} \epsilon_\pm^\nu(p) = \pm\epsilon_\pm^\mu(p)\ , \\
J_3 \epsilon_\pm^\mu(p) =& \mathcal{J}^{(3)\mu}_{\hspace{0.2in}\nu} \epsilon_0^\nu(p) = 0 \ ,\\
J_- \epsilon_+^\mu(p) =& \mathcal{J}^{(-)\mu}_{\hspace{0.25in}\nu} \epsilon_+^\nu(p) = \sqrt{2}\epsilon_0^\mu(p)\ ,\\
J_- \epsilon_0^\mu(p) =& \mathcal{J}^{(-)\mu}_{\hspace{0.25in}\nu} \epsilon_0^\nu(p) = \sqrt{2}\epsilon_-^\mu(p)\ ,\\
J_- \epsilon_-^\mu(p) =& \mathcal{J}^{(-)\mu}_{\hspace{0.25in}\nu} \epsilon_-^\nu(p) = 0 \ ,\\
J_+ \epsilon_-^\mu(p) =& \mathcal{J}^{(+)\mu}_{\hspace{0.25in}\nu} \epsilon_-^\nu(p) = \sqrt{2}\epsilon_0^\mu(p)\ ,\\
J_+ \epsilon_0^\mu(p) =& \mathcal{J}^{(+)\mu}_{\hspace{0.25in}\nu} \epsilon_0^\nu(p) = \sqrt{2}\epsilon_+^\mu(p)\ ,\\
J_+ \epsilon_+^\mu(p) =& \mathcal{J}^{(+)\mu}_{\hspace{0.25in}\nu} \epsilon_+^\nu(p) = 0\ ,
\esp\ee
so that we observe that no additional phase is required and the particle spin raising and lowering operators satisfy
\begin{equation}
J^1_\pm \epsilon_\sigma^\mu(p) = \sqrt{(1\mp\sigma)(1\pm\sigma+1)}\ \epsilon_{\sigma\pm1}^\mu(p) \ .
\end{equation}

For  antiparticles, we find
\be\bsp
\bar{J}_3 \epsilon_\pm^{\mu*}(p) =& -\mathcal{J}^{(3)\mu}_{\hspace{0.2in}\nu} \epsilon_\pm^{\nu*}(p) = \pm\epsilon_\pm^{\mu*}(p)\ ,  \\
\bar{J}_3 \epsilon_\pm^{\mu*}(p) =& -\mathcal{J}^{(3)\mu}_{\hspace{0.2in}\nu} \epsilon_0^{\nu*}(p) = 0 \ ,\\
\bar{J}_- \epsilon_+^{\mu*}(p) =& -\mathcal{J}^{(+)\mu}_{\hspace{0.25in}\nu} \epsilon_+^{\nu*}(p) = \sqrt{2}\epsilon_0^{\mu*}(p)\ , \\
\bar{J}_- \epsilon_0^{\mu*}(p) =& -\mathcal{J}^{(+)\mu}_{\hspace{0.25in}\nu} \epsilon_0^{\nu*}(p) = \sqrt{2}\epsilon_-^{\mu*}(p)\ , \\
\bar{J}_- \epsilon_-^{\mu*}(p) =& -\mathcal{J}^{(+)\mu}_{\hspace{0.25in}\nu} \epsilon_-^{\nu*}(p) = 0 \ ,\\
\bar{J}_+ \epsilon_-^{\mu*}(p) =& -\mathcal{J}^{(-)\mu}_{\hspace{0.25in}\nu} \epsilon_-^{\nu*}(p) = \sqrt{2}\epsilon_0^{\mu*}(p)\ , \\
\bar{J}_+ \epsilon_0^{\mu*}(p) =& -\mathcal{J}^{(-)\mu}_{\hspace{0.25in}\nu} \epsilon_0^{\nu*}(p) = \sqrt{2}\epsilon_+^{\mu*}(p)\ , \\
\bar{J}_+ \epsilon_+^{\mu*}(p) =& -\mathcal{J}^{(-)\mu}_{\hspace{0.25in}\nu} \epsilon_+^{\nu*}(p) = 0\ ,
\esp\ee
which again shows us there is no extra phase associated with these polarization vectors.  The raising and lowering operators satisfy in this case
\begin{equation}
\bar{J}^1_\pm \epsilon_\sigma^{\mu*}(p) = \sqrt{(1\mp\sigma)(1\pm\sigma+1)}\ \epsilon_{\sigma\pm1}^{\mu*}(p) \ .
\end{equation}

\subsubsection{Spin-$\tha$ fields}
The spin-$\tha$ polarization vectors can be formed as a direct product of the spin-$\oh$ and spin-$1$ polarization vectors.
The generators are then simply the sum of those of the spin-$\oh$ and spin-$1$ representations,
\be
J^{3/2}_3= J^{1/2}_3 + J^1_3 
\quad\text{and}\quad
J^{3/2}_\pm = J^{1/2}_\pm + J^1_\pm \ .
\ee
It is easily seen that these satisfy the $\mathfrak{su}(2)$ algebra.

We begin with the highest weight of the spin-$\tha$ representation which has helicity $\tha$ and is given by the product of the highest polarization of spin-$\oh$ and spin-$1$
\begin{equation}
u^\mu_{3/2,3/2}(p) = \epsilon^{\mu}_{+}(p)u_{+1/2}(p) \ .
\end{equation}
We then lower the $J_3$-eigenvalue by
using the $J^{3/2}_-$ operator to get the other states, $J^{1/2}_-$ only acting
on spin-$\oh$ polarization vectors and $J^1_-$ on spin-$1$ polarization vectors.  We now have the choice of the overall phase for the raising and lowering operators.  We take the trivial choice with no additional phase,
\begin{equation}
J^{3/2}_\pm u^\mu_{3/2,\sigma}(p) = \sqrt{\left(\tha\mp\sigma\right)\left(\tha\pm\sigma+1\right)}\ u^\mu_{3/2,\sigma\pm1}(p) \ .
\end{equation}
We now work out the first step explicitly,
\be\bsp
 &J^{3/2}_- u^\mu_{3/2,3/2}(p) = \sqrt{3} u^\mu_{3/2,1/2}(p)\\
&\quad = \left(J^1_-\epsilon^\mu_+(p)\right)u_{+1/2}(p)+\epsilon^\mu_+\left(J^{1/2}_- u_{+1/2}(p)\right)\\
&\quad = \sqrt{2}\epsilon^\mu_0(p)u_{+1/2}(p)+e^{i\phi}\epsilon^\mu_+(p)u_{-1/2}(p)\ ,
\esp\ee
so that
\begin{equation}
u^\mu_{3/2,1/2}(p) = \sqrt{\frac{2}{3}}\epsilon^\mu_0(p)u_{+1/2}(p)+\frac{1}{\sqrt{3}}e^{i\phi}\epsilon^\mu_+(p)u_{-1/2}(p) \ .
\end{equation}
Following the same procedure for the other states gives
\be\bsp
u^\mu_{3/2,-1/2}(p) =&\ \sqrt{\frac{2}{3}}e^{i\phi}\epsilon^\mu_0(p)u_{-1/2}(p) \\ &\quad 
   +\frac{1}{\sqrt{3}}\epsilon^\mu_-(p)u_{+1/2}(p) \ , \\
u^\mu_{3/2,-3/2}(p) =&\ e^{i\phi}\epsilon^{\mu}_{-}(p)u_{-1/2}(p) \ .
\esp\ee

Antiparticle polarization vectors are derived in a similar fashion using the conjugate lowering operators and
starting with the highest weight of the antiparticle polarization vector,
\be\bsp
v^\mu_{3/2,3/2}(p) =&\ \epsilon^{\mu *}_{+}(p)v_{+1/2}(p)\ ,\\
v^\mu_{3/2,1/2}(p) =&\ \sqrt{\frac{2}{3}}\epsilon^{\mu
  *}_0(p)v_{+1/2}(p)\\ &\quad +\frac{1}{\sqrt{3}}e^{-i\phi}\epsilon^{\mu *}_+(p)v_{-1/2}(p)\ ,\\
v^\mu_{3/2,-1/2}(p) =&\ \sqrt{\frac{2}{3}}e^{-i\phi}\epsilon^{\mu
  *}_0(p)v_{-1/2}(p)\\ &\quad +\frac{1}{\sqrt{3}}\epsilon^{\mu *}_-(p)v_{+1/2}(p)\ ,\\
v^\mu_{3/2,-3/2}(p) =&\ e^{-i\phi}\epsilon^{\mu *}_{-}(p)v_{-1/2}(p) \ ,
\esp\ee
where we have again not included any extra phase in the raising and lowering operators.

\subsubsection{Spin-$2$ fields}
Just as in the case of the spin-$\tha$ fields, the spin-$2$ polarization vectors can be formed as the direct product of two spin-$1$ polarization vectors.
The generators are hence the sum of the spin-$1$ generators for each polarization vector,
\be
J^{2}_3= J^1_3 + J^1_3 
\quad\text{and}\quad
J^{2}_\pm = J^1_\pm + J^1_\pm \ ,
\ee
where it is understood that the first generator only acts on the first polarization vector and the second generator only acts on the second polarization vector.
It is easily seen that these satisfy the $\mathfrak{su}(2)$ algebra.

We begin with the highest weight of the spin-$2$ representation which has helicity $2$ and is given by the product of the highest polarization of the two spin-$1$ polarization vectors,
\begin{equation}
\epsilon^{\mu\nu}_{2,2}(p) = \epsilon^{\mu}_{+}(p)\epsilon^{\nu}_{+}(p) \ .
\end{equation}
We then lower the $J_3$-eigenvalue by
using the $J^{2}_-$ operator to get the other states.  
Again, we choose to not include an additional phase,
\begin{equation}
J^{2}_\pm \epsilon^{\mu\nu}_{2,\sigma}(p) = \sqrt{\left(2\mp\sigma\right)\left(2\pm\sigma+1\right)}\ \epsilon^{\mu\nu}_{2,\sigma\pm1}(p) \ .
\end{equation}
We now work out the first step explicitly,
\be\bsp
 &J^{2}_- \epsilon^{\mu\nu}_{2,2}(p) = 2 \epsilon^{\mu\nu}_{2,1}(p)\\
&\quad = \left(J^1_-\epsilon^\mu_+(p)\right)\epsilon^\nu_{+}(p)+\epsilon^\mu_+\left(J^{1}_- \epsilon^\nu_{+}(p)\right)\\
&\quad = \sqrt{2}\epsilon^\mu_0(p)\epsilon^\nu_{+}(p)+\sqrt{2}\epsilon^\mu_+(p)\epsilon^\nu_{0}(p)\ ,
\esp\ee
so that
\begin{equation}
\epsilon^{\mu\nu}_{2,1}(p) = \frac{1}{\sqrt{2}}\left(\epsilon^\mu_0(p)\epsilon^\nu_{+}(p)+\epsilon^\mu_+(p)\epsilon^\nu_0(p)\right) \ .
\end{equation}
Following the same procedure for the other states gives
\be\bsp
\epsilon^{\mu\nu}_{2,0}(p) =&\ \frac{1}{\sqrt{6}}\big(\epsilon^\mu_{-}(p)\epsilon^\nu_+(p)+2\epsilon^\mu_0(p)\epsilon^\nu_0(p)\\
   &\quad +\epsilon^\mu_+(p)\epsilon^\nu_-(p)\big) \ , \\
\epsilon^{\mu\nu}_{2,-1}(p) =&\ \frac{1}{\sqrt{2}}\big(\epsilon^\mu_-(p)\epsilon^\nu_0(p)+\epsilon^\mu_0(p)\epsilon^\nu_-(p)\big) \ , \\
\epsilon^{\mu\nu}_{2,-2}(p) =&\ \epsilon^{\mu}_{-}(p)\epsilon^\nu_{-}(p) \ .
\esp\ee

Antiparticle polarization vectors are derived in a similar fashion using the conjugate lowering operators and
starting with the highest weight of the antiparticle polarization vector, which gives
\begin{equation}
\bar{\epsilon}^{\mu\nu}_{2,\sigma}(p) = \epsilon^{\mu\nu *}_{2,\sigma}(p) \ .
\end{equation}
We have again not included any extra phase in the raising and lowering operators.

\subsection{\label{sec:propagator}Propagators}
On-shell, a propagator is fully determined by the spin of the particle.  In particular, the numerator of the propagator is given by the sum over the polarization vectors in the following way
\begin{equation}
\lim_{p^2\to M^2}(p^2-M^2+i\epsilon)\Delta^j(p) = \sum_{\sigma} u_{j,\sigma}(p) u^*_{j,\sigma}(p) \ ,
\end{equation}
where $\Delta^j(p)$ is the full propagator for a spin-$j$ particle.
Off-shell, the propagator can be different as long as the difference vanishes when evaluated on-shell.
This difference is due to the presence of lower spin components in the propagator which are removed on-shell by the classical equations of motion for the field.
These extra components are determined by the associated terms in the quadratic part of the Lagrangian,
which is model-dependent. In this section, we show that the on-shell relation above holds in the
spin-$\oh$, spin-$1$, spin-$\tha$, and spin-$2$ cases.

\subsubsection{Spin-$\oh$ fields}
Using the polarization vectors given in Section~\ref{sec: spin 1/2 polarization vectors}, we find
\begin{eqnarray}
\sum_{\sigma=-1/2}^{1/2} u_\sigma(p)\bar{u}_\sigma(p) = &&
\end{eqnarray}
\begin{equation}
\left(\begin{array}{cccc}
M & 0 & E-|\vec{p}|\cos\theta & -e^{-i\phi}|\vec{p}|\sin\theta\\
0 & M & -e^{i\phi}|\vec{p}|\sin\theta & E+|\vec{p}|\cos\theta\\
E+|\vec{p}|\cos\theta & e^{-i\phi}|\vec{p}|\sin\theta & M & 0\\
e^{i\phi}|\vec{p}|\sin\theta & E-|\vec{p}|\cos\theta & 0 & M
\end{array}\right)\nonumber \ .
\end{equation}
This is to be compared with the spin-$\oh$ propagator, which is exactly the same both on-shell and off-shell
since there is no possible spin lower by an integer that can contribute,
\begin{equation}
\slashed{p}+M = \sum_{\sigma=-1/2}^{1/2} u_\sigma(p)\bar{u}_\sigma(p) \ .
\end{equation}
So, the spin-$\oh$ propagator is fixed entirely by its spin.

We have also checked that antiparticle polarization vectors satisfy a similar relation,
\begin{equation}
\slashed{p}-M = \sum_{\sigma=-1/2}^{1/2} v_\sigma(p)\bar{v}_\sigma(p) \ .
\end{equation}

\subsubsection{Spin-1 fields}
Again, using the particle polarization vectors given in Section~\ref{sec: spin 1 polarization vectors}, we find
\begin{equation}
\sum_{\sigma=-1}^1 \epsilon_\sigma^\mu\epsilon_\sigma^{\nu*} = 
\frac{1}{M^2}\left(\begin{array}{cc}
|\vec{p}|^2 & \cdots \\
E|\vec{p}|\sin\theta\cos\phi & \cdots \\
E|\vec{p}|\sin\theta\sin\phi & \cdots \\
E|\vec{p}|\cos\theta & \cdots
\end{array}\right) \ .
\end{equation}
The propagator numerator is given by
\begin{equation}
\Pi_1^{\mu\nu}=-\eta^{\mu\nu}+\frac{p^\mu p^\nu}{M^2} = 
\frac{1}{M^2}\left(\begin{array}{cc}
E^2-M^2 & \cdots \\
E|\vec{p}|\sin\theta\cos\phi & \cdots \\
E|\vec{p}|\sin\theta\sin\phi & \cdots \\
E|\vec{p}|\cos\theta & \cdots
\end{array}\right) \ .
\end{equation}
In these last two equations, we have omitted the matrix elements of the second, third and
fourth column for brevity.
We have however checked that they all agree with each other on-shell, so that
\begin{equation}
\lim_{p^2\to M^2} \left(-\eta^{\mu\nu}+\frac{p^\mu p^\nu}{M^2}\right) = \sum_{\sigma=-1}^1 \epsilon_\sigma^\mu\epsilon_\sigma^{\nu*} \ .
\end{equation}
Similar results can be obtained starting from antiparticle polarization vectors.

In the off-shell case, we find that the propagator numerator differs from the sum over the polarization vectors
due to the presence of spin-$0$ components controlled by the quadratic part of the Lagrangian.  It can
be noted that for spin-$1$ fields, unitarity also plays a role in fixing the propagator.

\subsubsection{Spin-$\tha$ fields}
In this case, there are four sets of indices so that we omit to display the entire results in this paper
for brevity.  However, we have checked every element and find that 
\begin{equation}
\lim_{p^2\to M^2}\Pi_{RS}^{\mu\nu} 
  = \sum_{i=-3/2}^{3/2} u_{3/2,i}^\mu \bar{u}_{3/2,i}^\nu \ ,
\end{equation}
where the numerator of the propagator reads
\be\bsp
\Pi_{RS}^{\mu\nu} =& \
  \Big[-\eta^{\mu\nu} + \frac{p^\mu p^\nu}{M^2}\Big] \Big[ \slashed{p} + M\Big]\\
  &\quad - \frac13 \Big[\gamma^\mu + \frac{p^\mu}{M} \Big]
 \Big[\slashed{p}-M\Big] \Big[ \gamma^\nu + \frac{p^\nu}{M}\Big] \\
 =&\
 -\Big[\slashed{p}+M\Big]\Big[ \eta^{\mu\nu} -  \frac23 \frac{p^\mu p^\nu}{M^2}
   -   \frac13 \gamma^\mu\gamma^\nu 
 \\ &\quad  - \frac{1}{3M}\big(p^\nu\gamma^\mu - p^\mu\gamma^\nu\big) \Big] \ .
\esp\label{eq:sp32prop}\ee
This is the propagator used by both {\sc CalcHEP}
and {\sc MadGraph}~5.
We have also checked that the related antiparticle relation
\begin{equation}
\lim_{p^2\to M^2} \Pi_{RSanti}^{\mu\nu} = \sum_{i=-3/2}^{3/2} v_{3/2,i}^\mu \bar{v}_{3/2,i}^\nu \ ,
\end{equation}
holds, where
\be\bsp
\Pi_{RSanti}^{\mu\nu} =& \
  \Big[-\eta^{\mu\nu} + \frac{p^\mu p^\nu}{M^2}\Big] \Big[ \slashed{p} - M\Big]\\
  &\quad - \frac13 \Big[\gamma^\mu - \frac{p^\mu}{M} \Big]
 \Big[\slashed{p}+M\Big] \Big[ \gamma^\nu - \frac{p^\nu}{M}\Big] \\
 =&\
 -\Big[\slashed{p}-M\Big]\Big[ \eta^{\mu\nu} -  \frac23 \frac{p^\mu p^\nu}{M^2}
   -   \frac13 \gamma^\mu\gamma^\nu 
 \\ &\quad  + \frac{1}{3M}\big(p^\nu\gamma^\mu - p^\mu\gamma^\nu\big) \Big]\ .
\esp\ee
Many terms are different in the off-shell case due to the presence of spin-$\oh$ components.  
These are model-dependent, as is the exact form of the propagator.  For example, another spin-$\tha$
propagator that has been used in the literature \cite{Hagiwara:2010pi} has its numerator given by
\be\bsp
\tilde\Pi^{\mu\nu} = &\ \Big(\slashed{p}+M\Big)
\bigg[\left(-\eta^{\mu\nu}+\frac{p^\mu p^\nu}{M^2}\right)\\ &\quad +  
\frac{1}{3} \left(\eta^{\mu\alpha}-\frac{p^\mu p^\alpha}{M^2}
\right)\left( \eta^{\nu\beta}-\frac{p^\nu p^\beta}{M^2} \right)
\gamma_\alpha \gamma_\beta\bigg] \ .
\esp\label{eq:propRS2}\ee
The difference between the numerators of the propagators of Eq.~\eqref{eq:sp32prop} and Eq.~\eqref{eq:propRS2} is
\be
\Pi_{RS}^{\mu\nu}\!-\!\tilde\Pi^{\mu\nu} \!=\! \frac{(p^2\!-\!M^2)}{3M^2}\bigg[\frac{p^\mu p^\nu}{M^2}(\slashed{p}
  \!+\! M)\!-\!(p^\mu\gamma^\nu\!-\!p^\nu\gamma^\mu)\bigg] \ ,
\ee
which vanishes on-shell.

\subsubsection{Spin-$2$ fields}\label{sec:spin2prop}
For spin-$2$, one form of the propagator numerator is given by
\begin{equation}
\Pi_2^{\mu\nu\alpha\beta}=\frac{1}{2}\Pi_1^{\mu\alpha}\Pi_1^{\nu\beta}+\frac{1}{2}\Pi_1^{\mu\beta}\Pi_1^{\nu\alpha}-\frac{1}{3}\Pi_1^{\mu\nu}\Pi_1^{\alpha\beta} \ ,
\end{equation}
which is the propagator used by both {\sc CalcHEP}
and {\sc MadGraph}~5.  Other forms differ in their off-shell effects.
The resulting tensor is much too long for every term to be included.  We have however
checked every term and find exact agreement with the sum of the polarization vectors as in
\begin{equation}
\lim_{p^2\to M^2}\Pi_2^{\mu\nu\alpha\beta}=\sum_{\sigma=-2}^{2}\epsilon_{2,\sigma}^{\mu\nu}\epsilon_{2,\sigma}^{\alpha\beta *} \ .
\end{equation}

\subsection{\label{sec: Lagrangian}Quadratic Lagrangian}
In this subsection, we have followed the approach suggested by Weinberg, noting that the spin determines the on-shell propagator without any discussion of the Lagrangian.  So, in this way, we learn that the spin also fixes the on-shell quadratic terms of the Lagrangian.  The off-shell piece is model-dependent, as we have mentioned with the propagator being the inverse of the quadratic Lagrangian terms off-shell as well as on-shell. 

In this section, we show that the quadratic Lagrangian pieces correspond to
the inverse of the propagator for spin-$\oh$, spin-$1$, spin-$\tha$ and spin-$2$ fields.  

\subsubsection{Spin-$\oh$ fields}
In this case, there is no ambiguity for either the propagator or the quadratic Lagrangian which reads
\begin{equation}
{\cal L} = \bar{\psi}\left(\slashed{p}-M\right)\psi \ ,
\end{equation}
for a spin-$\oh$ fermionic field $\psi$. We can easily verify that this operator is the
inverse of the propagator, since
\begin{equation}
\lim_{\epsilon\to 0}\left[
\frac{\slashed{p}+M}{p^2-M^2+i\epsilon}\times \left(\slashed{p}-M\right)
\right] = 1 \ .
\end{equation}

\subsubsection{Spin-1 fields}
Although, in general, the spin-$1$ Lagrangian is ambiguous, the unitarity condition requires massive vector bosons to obtain their mass from spontaneous symmetry breaking.  Therefore, the Lagrangian for a vector boson field $V_\mu$
is required to be gauge invariant.  After spontaneous symmetry breaking, the quadratic piece is of the form
\begin{equation}
{\cal L} = \frac{1}{2}V^\mu\left(\left(-p^2+M^2\right)\eta_{\mu\nu}+p_\mu p_\nu\right)V^\nu \ ,
\end{equation}
where we have inserted a factor of $1/2$ since gauge bosons are real fields.  Multiplying this operator by the 
associated propagator gives
\be
\lim_{\epsilon\to 0}\bigg[
\frac{-\eta^{\mu\nu}+\frac{p^\mu p^\nu}{M^2}}{p^2-M^2+i\epsilon}\Big((-p^2+M^2)\eta_{\nu\omega}+p_\nu p_\omega\Big)
\bigg]
\!=\! \delta^\mu{}_\omega \ .
\ee
This proves the original statement in the spin-1 case.

\subsubsection{Spin-$\tha$ fields}
The Lagrangian describing the dynamics of a free spin-$\tha$ field  $\Psi_{\mu}$
of mass $M$ can be casted under several forms, all
widely used in the literature. The one associated with the equations of
motion presented in Eq.~\eqref{raritaschwingereq} read
\be
 {\cal L} = \epsilon^{\mu\nu\rho\sigma} \bar{\Psi}_\mu
   \gamma_5 \gamma_\sigma \partial_\nu\Psi_\rho
   +  2 i M \bar\Psi_\mu\gamma^{\mu\nu}\Psi_\nu  \ ,
\label{eq:lageps}\ee
our conventions on the Dirac matrices and antisymmetric tensors being collected in Appendix~\ref{sec:app:conventions}.
Equivalently, the original form introduced by Rarita and Schwinger \cite{Rarita:1941mf},
\be\bsp
 {\cal L} =&\ \bar\Psi_\mu\big(-i \slashed{\partial} + M\big)\Psi^\mu
 + i \bar\Psi_\mu \big( \gamma^\mu\partial^\nu + \gamma^\nu\partial^\mu\big)\Psi_\nu 
 \\ &\
 - i \bar\Psi_\mu \gamma^\mu\slashed{\partial}\gamma^\nu \Psi_\nu
  - M \bar\Psi_\mu\gamma^\mu\gamma^\nu\Psi_\nu\\
  \equiv &\ \bar\Psi_\mu \Lambda^{\mu\nu}\Psi_\nu \ ,
\esp\label{eq:lagori}\ee
can be retrieved after employing the property of the Dirac matrices of Eq.~\eqref{eq:Diracprop1},
together with the Clifford algebra relation of Eq.~\eqref{eq:cliff}.
It can be easily checked, after making use of Eq.~\eqref{eq:cliff}, that the
operator appearing in Eq.~\eqref{eq:lagori} is inverted by the propagator
of Eq.~\eqref{eq:sp32prop}, 
\be\bsp
  \lim_{\eps\to 0}\bigg[ \frac{\Pi^{\mu\nu}_{RS}}{p^2-M^2+i\epsilon} \Lambda_{\nu\rho} \bigg] =
    \delta^\mu{}_\rho \ .
\esp\ee

\subsubsection{Spin-$2$ fields}
The standard Lagrangian for a massive graviton field $h_{\mu\nu}$ with a mass $M$
is obtained starting from the Fierz-Pauli La\-gran\-gi\-an~\cite{Fierz:1939ix}. In
this case, it is given by
\be\bsp
  {\cal L} = &\
    \frac12 \partial_\rho h_{\mu\nu} \partial^\rho h^{\mu\nu} -
    \frac12 \partial_\rho h^\mu{}_\mu \partial^\rho h^\nu{}_\nu\\ &\ +
    \partial_\rho h^\mu{}_\mu \partial_\nu h^{\rho\nu}  -
    \partial_\rho h_{\mu\nu} \partial^\nu h^{\mu\rho} \\ &\ 
    -\frac{M^2}{2} \big(h_{\mu\nu} h^{\mu\nu} - h^\mu{}_\mu h^\nu{}_\nu \big) \ ,
\esp\ee
where the structure of the mass term follows the conventions of the so-called
Fierz-Pauli tuning \cite{Hinterbichler:2011tt} as usually employed to prevent ghost contributions
from being required. After integrating the previous Lagrangian by parts, it can be rewritten as
\be
  {\cal L} = \frac12 h_{\mu\nu} {\cal O}^{\mu\nu,\alpha\beta} h_{\alpha\beta} \ ,
\ee
where the operator ${\cal O}^{\mu\nu,\alpha\beta}$ is given by
\be\bsp
  {\cal O}^{\mu\nu,\alpha\beta} \! = &
   \frac12 \Big[\eta^{\mu\alpha}\eta^{\nu\beta} \!+\! \eta^{\mu\beta}\eta^{\nu\alpha}
     \!-\! 2 \eta^{\mu\nu} \eta^{\alpha\beta}\Big] \Big[p^2 \!-\! M^2\Big]
\\ &
 + p^\mu p^\nu \eta^{\alpha\beta}
 + p^\alpha p^\beta \eta^{\mu\nu}
 - \frac12 p^\mu p^\alpha \eta^{\nu\beta}
\\ & - \frac12 p^\mu p^\beta  \eta^{\nu\alpha}
 - \frac12 p^\nu p^\alpha \eta^{\mu\beta}
 - \frac12 p^\nu p^\beta  \eta^{\mu\alpha} \ .
%
%
\esp\ee
Under this form, it can be easily verified that this operator
is the inverse of the massive spin-two propagator defined in Section~\ref{sec:spin2prop},
\be
  \lim_{\eps\to 0}\bigg[ \frac{\Pi^{\mu\nu\alpha\beta}_2}{p^2-M^2+i\epsilon} {\cal O}_{\alpha\beta, \rho\sigma} \bigg] =
    \frac12 \Big[  \delta^\mu{}_\rho \delta^\nu{}_\sigma +  \delta^\mu{}_\sigma \delta^\nu{}_\rho \Big] \ .
\ee

\subsection{\label{sec:Wigner d-functions}Wigner $d$-functions}
Since total angular momentum is conserved in any system, if the
angular momentum of the initial state is described by a $| j,m \rangle$ state
with $m$ being along an initial direction, then so will the final state.
However, if we measure the final state along a different direction, we need
to calculate the corresponding wave function overlap between the two directions under consideration. 
We suppose that the final state is described by $|j,m',\theta\rangle$, where
$m'$ is the component of the angular momentum along the final
direction which makes an angle $\theta$ with respect to the initial
direction.  The wavefunction overlap between these two bases is given
by the product
\begin{equation} 
\langle j,m',\theta | j,m \rangle \ .
\end{equation}
These different bases are related by the rotation operator $\vec{J}$,
so that
\begin{equation}
| j,m',\theta\rangle = e^{-i\theta J_\perp} | j,m'\rangle \ ,
\end{equation}
where $J_\perp$ is the component of the rotation operator that is
perpendicular to the plane spanned by the initial direction and the final
direction. A rotation of angle $\theta$ hence rotates the
initial basis into the final basis.
As a result, the wavefunction overlap is given by
\begin{equation}
d^j_{m'm} (\theta) = \langle j, m',\theta |  j,m \rangle  = \langle j, m'| e^{i \theta J_\perp} | j,m \rangle\ .
\end{equation}
These functions of $\theta$ are called the Wigner $d$-functions.  Since
they only depend on the conservation of angular momentum, they can be
calculated independently of the internal dynamics defined by the
Lagrangian.   
For an illustrative derivation of the Wigner $d$-functions for $j=0,1$, we refer to
the Appendix C of Ref.~\cite{Chiang:2011kq}.  The full set of Wigner $d$-functions for
$j\leq2$ can be found in the Particle Data Group review~\cite{Beringer:2012zz}
and satisfy the relations
\be\label{reldj}\bsp
 d^j_{mm'} (\theta) =&\ (-1)^{m-m'} d^j_{m'm} (\theta) \\
=&\ d^j_{-m',-m}(\theta) \\
=&\ (-1)^{j-m} d^j_{m,-m'} (\pi-\theta) \ .
\esp\ee

Although the Wigner $d$-functions apply to any decay or collision, they are often
discussed in the context of a $2\to2$ process in the
center-of-momentum frame.  
For this case, we define the initial spin component direction to be along the
direction of motion of one of the initial particles.  Since the other
initial particle has momentum equal in size but opposite in direction,
we find that the initial spin component is equal to the difference of
the initial helicities.  We will define the final spin component
direction to be along the direction of motion of one of the final state
particles.  As in the initial particle case, we find that the final
spin component is equal to the difference of the helicities of the
final states.  Although the orbital angular momentum can contribute to
the total angular momentum $j$, since it is perpendicular to the plane
of the collision, it does not contribute to the component of either
the initial angular momentum or the final angular momentum.  So, if
the total angular momentum is $j$ (after including orbital angular
momentum), the initial difference of helicity is $m$, and the final
difference of helicity is $m'$, we find that the angular dependence of
the collision must be given by $d^j_{m,m'}(\theta)$.

One consequence of the conservation of angular momentum is that any
scattering amplitude can be expanded in terms of the Wigner $d$-functions (this
is called a partial wave expansion).
\begin{equation}
{\cal M}_{m'm} = \sum_{j} M_{jm'm} d^j_{m'm} (\theta) \ .
\label{Mamp}
\end{equation}
This equation only determines which Wigner $d$-functions are
allowed for a certain scattering process, independent of the internal
dynamics.  However, on the other hand, the values of the $M_{jmm'}$
are determined by a combination of the initial spins, the
final spins and the internal dynamics.  For example, an
electron-positron collider can be polarized so that the contribution
to $m=1$ is larger than the contribution to $m=-1$.  For another
example, if we know that at any moment in the scattering process,
there is only one particle in the $s$-channel and it is on its mass shell, then we know
that the total angular momentum must equal the spin of that particle.
Since there is only one particle, it is at rest in the collision center-of-momentum
frame, so there is no orbital angular momentum at that moment. This
means that only the $M_{jm'm}$ with $j$ equal to the spin of this
particle are non-zero,
\begin{equation}
{\cal M}_{m'm} = M_{jm'm} d^j_{m'm} (\theta) \ .
\label{Mamp}
\end{equation}

%% file: helgrav.tex
In this Appendix, we present analytical expressions for the helicity amplitudes
${\cal M}_{\lam_1\lam_2,\lam_3\lam_4}$
associated with the process
\be
 g(p_1,\lam_1)+g(p_2,\lam_2)\to\gld(p_3,\lam_3)+\gld(p_4,\lam_4)\ ,
\ee
where the four-momentum ($p_i$) and the helicity ($\lam_{1,2}=\pm 1$ and
$\lam_{3,4}=\pm \tha,\pm \oh$) are explicitly indicated.
The helicity amplitudes can be expressed as a
sum upon $s$-, $t$- and $u$-channel contributions,
\be
 {\cal M}_{\lam_1\lam_2,\lam_3\lam_4}={\cal M}^s+{\cal M}^t+{\cal M}^u,
\ee
with the helicity dependence of the right-hand side of the equation
being understood for clarity. We derive, starting from
the Feynman rules extracted from the Lagrangians of Eq.~\eqref{lag_grv},
\be\bsp
 i{\cal M}^s &\ =-\frac{1}{\Mpl^2}\frac{1}{s}\,
   \epsilon_{\mu}(p_1)\epsilon_{\nu}(p_2) \\
   \times&\ \Gamma_g^{\mu\nu,\alpha\beta} 
   B_{\alpha\beta,\gamma\delta}\,
   \bar u_{\rho}(p_3)\Gamma_{\gld}^{\gamma\delta,\rho\sigma}
   v_{\sigma}(p_4)\ , \\
 i{\cal M}^t &\ =\frac{i}{16\Mpl^2}\frac{1}{t_{\go}}\,
   \epsilon_{\mu}(p_1)\epsilon_{\nu}(p_2) \\
   \times&\ \bar{u}_{\rho}(p_3)[\slashed{p}_1,\gamma^{\mu}]\gamma^{\rho}
   (\slashed{p}_3-\slashed{p}_1+m_{\go})
   \gamma^{\sigma}[\gamma^{\nu},\slashed{p}_2]v_{\sigma}(p_4)\ , \\
 i{\cal M}^u &\ =\frac{i}{16\Mpl^2}\frac{1}{u_{\go}}\,
   \epsilon_{\mu}(p_1)\epsilon_{\nu}(p_2) \\
   \times&\ \bar{u}_{\rho}(p_3)[\slashed{p}_2,\gamma^{\nu}]\gamma^{\rho}
   (\slashed{p}_1-\slashed{p}_4+m_{\go})
   \gamma^{\sigma}[\gamma^{\mu},\slashed{p}_1]v_{\sigma}(p_4)\ , 
\esp\ee
where the Mandelstam variables are defined by
$s=(p_1+p_2)^2$, $t_{\go}=(p_1-p_3)^2-m_{\go}^2$ and $u_{\go}=(p_1-p_4)^2-m_{\go}^2$,
$m_{\go}$ being the gluino mass.
The Lorentz structure of each amplitude has been embedded into the functions
\be\bsp
 B^{\alpha\beta,\gamma\delta}&\ =\frac{1}{2}(\eta^{\alpha\gamma}\eta^{\beta\delta}+\eta^{\alpha\delta}\eta^{\beta\gamma}-\eta^{\alpha\beta}\eta^{\gamma\delta}) \ , \\
 \Gamma_g^{\mu\nu,\alpha\beta}
 &\ = (p_1\cdot p_2)\,C_{\alpha\beta,\mu\nu} 
   +D_{\alpha\beta,\mu\nu}+E_{\alpha\beta,\mu\nu}\ ,\\
 \Gamma_{\gld}^{\gamma\delta,\rho\sigma}
 &\ =\frac{1}{4}\epsilon^{\rho\sigma\lambda\gamma}\gamma^5\gamma^{\delta}(p_3-p_4)_{\lambda}+(\gamma\leftrightarrow\delta)\\
&\ +\frac{i}{8}\epsilon^{\rho\sigma\lambda\gamma}\gamma^5\{\gamma_{\lambda},\sigma^{\delta\tau}\}(p_3+p_4)_{\tau}+(\gamma\leftrightarrow\delta)\\
&\ -im_{3/2}(\eta^{\rho\gamma}\eta^{\delta\sigma}+\eta^{\sigma\gamma}\eta^{\delta\rho}-\eta^{\gamma\delta}\eta^{\rho\sigma})\ ,
\esp\ee
with
\be\bsp
 C^{\alpha\beta,\mu\nu}
 &\ =\eta^{\alpha\mu}\eta^{\beta\nu}+\eta^{\alpha\nu}\eta^{\beta\mu}-\eta^{\alpha\beta}\eta^{\mu\nu}\ ,\\ 
 D^{\alpha\beta,\mu\nu}
 &\ =p_{1}^{\nu}p_{2}^{\mu}\eta^{\alpha\beta} \\
 &\ +(p_{1}^{\alpha}p_{2}^{\beta}\eta^{\mu\nu}\!-\!p_{1}^{\nu}p_{2}^{\beta}\eta^{\alpha\mu}
  \!-\! p_{1}^{\beta}p_{2}^{\mu}\eta^{\alpha\nu} \!+\! (\alpha\leftrightarrow\beta))\ ,\\
 E^{\alpha\beta,\mu\nu}
 &\ =\eta^{\alpha\beta}(p_1^{\mu}p_1^{\nu}+p_2^{\mu}p_2^{\nu}+p_1^{\mu}p_2^{\nu})\\
&\ -(p_1^{\mu}p_1^{\beta}\eta^{\nu\alpha}+p_2^{\nu}p_2^{\beta}\eta^{\mu\alpha}+(\alpha\leftrightarrow\beta))\ .
\esp\ee